\documentclass[12pt]{article}
\usepackage{amsmath,amssymb}
\usepackage{hyperref,cite}
\usepackage{appendix}
\usepackage{graphicx,subfigure}
\newcommand{\ul}[1]{\underline{#1}}
\def\eq#1{{Eq.~(\ref{#1})}}
\def\fig#1{{Fig.~\ref{#1}}}
\newcommand{\Lag}{\mathcal{L}}
\renewcommand{\v}[1]{\ensuremath{\mathbf{#1}}}

\begin{document}
\title{{\bf Color-neutral heavy particle production in nucleus-nucleus collisions in the quasi-classical approximation}}
\author{
{\bf Tseh Liou}
\\[1cm] {\it\small Department of Physics, Columbia University, New York, New York 10027}}

\date{}

\maketitle

\thispagestyle{empty}

\begin{abstract}
  We use a diagrammatic approach to study color-neutral heavy particle production in nucleus-nucleus collisions in a quasi-classical approximation without small-$x$ evolution. In order to treat the two nuclei symmetrically, we use the Coulomb gauge which gives the appropriate light cone gauge for each nucleus. The resulting cross section is factorized into a product of two Weizs\"{a}cker-Williams gluon distributions of the two nuclei when the transverse momentum of the produced scalar particle is around the saturation momentum. We confirm our results in covariant gauge where the transverse momentum broadening of hard gluons can be described as a diffusion process. The transverse momentum factorization manifests itself in light cone gauge but not so clearly in covariant gauge. 
\end{abstract}

\thispagestyle{empty}

\newpage

\setcounter{page}{1}
\section{Introduction}
Transverse momentum dependent (TMD) parton distributions are necessary to study less inclusive processes in high energy heavy ion collisions \cite{Gribov}. TMD parton distributions contain more information about the parton small-$x$ degrees of freedom than the integrated parton distributions and require a new factorization formalism which is different from the traditional collinear factorization \cite{Collins}. TMD's have attracted much interests in recent years \cite{Kharzeev,Belitsky}. There are two different unintegrated gluon distributions usually encountered in the literature \cite{Kharzeev,universality}. The conventional unintegrated gluon distribution is related to the color dipole scattering amplitude, and most known processes are related to it \cite{universality}. Another unintegrated gluon distribution is constructed by solving classical Yang-Mills equations of motion in the McLerran-Venugopalan model, a simple model used to study the small-$x$ gluon distribution in the saturation region for large nuclei \cite{MVmodel,CGC}. In a light cone gauge, the valence quarks of a nucleus are treated as static color sources, and the gluon fields of the sources add coherently then become the non-Abelian Weizs\"{a}cker-Williams (WW) $A_{\mu}^{WW}$ field of the nucleus. The WW gluon distribution is quite different from the conventional unintegrated gluon and a few physical processes which can directly probe this gluon distribution are known \cite{universality}. 

In this paper, we study nucleus-nucleus collisions analytically. We always assume the strong coupling constant $\alpha\ll 1$, but take $\alpha^2 A^{1/3}$, with $A$ the atomic number of a nucleus, to be a large parameter. We further assume that there is no QCD evolution in the gluon distributions of nucleons in the nucleus. Instead of considering difficult processes, such as gluon production \cite{gluonAA,Gelis,Blaizot}, whose final state interactions are very complicated, we will limit ourselves to a much simpler case where the produced particle is color-neutral. We always assume that the mass of the scalar particle, $M$, is much larger than the saturation momentum $Q_{s}$. The effective interaction we will use is $\Lag_{eff}=-\frac{1}{4}g_{\phi}\phi F^{a}_{\mu\nu}F^{a\mu\nu}$, where $\phi$ is a scalar field. This effective interaction is used to studied Higgs production in pp and pA \cite{Lipatov,pp,higgs,Hautmann}, however, the phenomenological aspect is not our major interest in this paper. The advantage of using such an effective interaction in a study of nuclei collisions is that we do not have to worry about the final state interactions of the produced particle, all the effects are from the initial state of the nuclei. Therefore, we can view the process as follows: the gluon fields in the two nucleons are developed independently and gluon distributions are formed within each nucleus before the collision. During the collision the gluon fields are released by the nuclei and the scalar particle is produced. After the collision the nuclei have to rebuild their gluon fields which would induce further gluon radiation but it is irrelevant to the current problem because the particle is color neutral and knows nothing about the rebuilding processes. This simple physical picture is exactly what we have in the Color Glass Condensate framework, and the transverse momentum distribution that manifests itself in this process is the WW gluon distribution, which comes from neglecting the rebuilding processes of the gluon field of the nucleus. In a light cone (LC) gauge , the cross section, \eq{eq:csfull}, is factorized into a production of two unintegrated gluon distributions \eq{eq:unint} and involves a new gluon distribution \eq{eq:linear}, the so-called linear polarized gluon distribution \cite{linearlypolarized}. In different kinematic regimes the cross section takes different forms. When $\ul{l}^2\sim Q_{s}^2$, where $\ul{l}$ is the transverse momentum of the scalar particle, the cross section becomes a product of two WW gluon distributions of the two nuclei in coordinate space \eq{eq:cs}. Therefore, we also manage to find a process that can probe the WW gluon distribution in addition to the dijet production discussed in \cite{universality}. If $\ul{l}^{2}\gg Q_{s}^2$, the cross section is factorizable as given by \eq{eq:csspin}.

The major question in the study of nuclear collisions is deciding what gauge to use. Gluon distributions in the two fast moving nuclei cannot be described naturally in one single LC gauge. In order to treat the two nuclei symmetrically we use Coulomb gauge which is the essential ingredient of our calculation. The advantage of using Coulomb gauge is that the Coulomb gauge propagator connected to a object with a big plus momentum component is equivalent to the $A_{+}=0$ gauge propagator. Similarly for a left moving object Coulomb gauge is equivalent to $A_{-}=0$ gauge \cite{Jaroszewicz,overview}. It is not surprising that one can do the calculation in this mixture of gauges. After all the $A_{\mu}^{WW}$ field has only transverse components and is obtained from the classical equation of motion in $A_{-}=A_{+}=0$ gauge \cite{WWfield}. However, there are some technical subtleties behind this naive connection which will be discussed in detail in Section \ref{sec:gauge}. 

Instead of trying to solve the classical field equations of the two colliding nuclei, in Section \ref{sec:calculation} we will use a diagrammatic approach to study color-neutral particle production in the nucleus-nucleus collisions. In order to simplify the diagrams the Slavnov-Taylor-Ward (STW) identities \cite{wardidentities} will be heavily used. The diagrammatic STW identities were used in various contexts, for example, in obtaining a non-linear gluon evolution equation \cite{recombination}, in studying the quantum structure of WW fields \cite{gaugerotation}, and in a neutral current DIS off a large nucleus \cite{pA}. We will start with some low order examples, then illustrate how to generalize from only one nucleon in each nucleus to an arbitrary number of nucleons in each nucleus and how the contribution of all soft gluons are resummed by $A_{\mu}^{WW}$. The cross section is manifestly factorizable in the light cone calculation. In Section \ref{sec:covariant} we also perform a calculation in covariant gauge as a confirmation of the result we have in the LC calculation. The physical picture is slightly different in covariant gauge. A hard gluon released by a nucleus will be multiply scattered, both elastically and inelastically, by nucleons of the other nucleus as the two nuclei pass through each other. The transverse momentum distribution of the hard gluon will be gradually broadened by the multiple scattering and can be found by solving a diffusion equation \cite{diffusion,pA}. If we sum up all possible nucleons from which the hard gluon could come from in a nucleus the WW gluon distribution shows up right before the collision, however the cross section does not manifestly factorizable at first glance. This is additional evidence that the WW gluon distribution is the right type of gluon distribution for this process. Moreover, comparing the calculation from these two different approaches, we believe that transverse momentum factorization can be easily achieved in light cone gauge, or more precisely in Coulomb gauge. Therefore light cone gauge with appropriate regularization in the propagator and the Ward identities are the two essential ingredients in obtaining a transverse momentum factorized formula. To our knowledge this is the first process where an exact analytic formula has been formed for a physical process, involving momenta on the order of $Q_{s}$, in nucleus-nucleus collisions in the quasi-classical approximation. We might hope that this method can be further applied to study more complicated processes, such as gluon production.

\section{Coulomb and Light Cone Gauge}
\label{sec:gauge}
In the LC gauge $A\cdot n=0$, $n^2=0$, the Yang-Mills field propagator contains a singular denominator $1/(k\cdot n)$ and must be regularized. The propagator that will be used in the following calculation takes the form
\begin{equation}
  \label{eq:lightcone}
  D_{\alpha\beta}(k)=\frac{-i}{k^{2}+i\epsilon}\bigg(g_{\mu\nu}-\frac{n_{\alpha}k_{\beta}}{n\cdot k+i\epsilon}-\frac{n_{\beta}k_{\alpha}}{n\cdot k-i\epsilon}\bigg),
\end{equation}
where the momentum $k$ flows from $\alpha$ to $\beta$. Note that, due to the opposite direction of the momentum flow, the second and the third terms in the propagator always have the opposite $i\epsilon$ pole in the complex $k\cdot n$-plane. This way of regularizing the propagator was previously derived in canonical quantization \cite{Antonov} and in path integral formalism \cite{SlaFro} and more recently in a gauge transformation analysis \cite{Belitsky}. Moreover, it has been realized that the $i\epsilon$ prescriptions in the LC propagator are closely related to initial and final state interactions \cite{recombination,pA,saturation,Belitsky}. That is the $i\epsilon$ poles in the light cone propagator dictate how gluons propagate in the $x_{-}$- or $x_{+}$-directions when they are emitted from physical objects, like nucleons. For example, in $0=n\cdot A=A_{+}$ gauge, the propagator (\ref{eq:lightcone}) tells us that the gluon fields coming from a right-moving nucleon only propagate in the positive $x_{-}$-direction, or in terms of gauge rotation, gluons coming from a source with coordinate $x_{-}^{0}$ only gauge rotates other sources that satisfy $x_{-}>x_{-}^{0}$. With gluons only propagating in one certain direction the $i\epsilon$ prescriptions can be used to avoid initial or final state effects. In a neutral current DIS off a large nucleus the regularization can be chosen such that it can eliminate final state interactions \cite{pA}. Implementing different prescriptions, one can also shift a final state interaction to an initial state interaction \cite{Belitsky}.

However, \eq{eq:lightcone} cannot be used immediately in the calculation of nucleus-nucleus collisions, because one single LC gauge choice would make the two nuclei asymmetric. For example, if we choose $A_{+}=0$ gauge, then a right moving nucleus having a big plus momentum component naturally fits in with such a gauge choice. However, for a left moving nucleus it is awkward to develop gluon fields in such a gauge. The same problem arises if we want to use $A_{-}=0$ gauge. Therefore one single LC gauge, either $A_{+}=0$ or $A_{-}=0$, will make the two fast moving nuclei asymmetrical. Naturally one wants to use $A_{+}=0$ gauge for the right-moving nucleus while $A_{-}=0$ gauge for the left-moving nucleus, so that gluon fields can be developed naturally in each nucleus. Indeed, the gauge choice satisfying this need exists, although it is not obvious at first sight. It is the Coulomb gauge $\nabla\cdot \v{A}=0$ \cite{overview}, in which the propagator reads
\begin{equation}
  \label{eq:coulomb}
  D_{\alpha\beta}(k)=-\frac{i}{k^2}\bigg[g_{\alpha\beta}-\frac{N\cdot k(N_{\alpha}k_{\beta}+N_{\beta}k_{\alpha})-k_{\alpha}k_{\beta}}{\v{k}^2}\bigg],
\end{equation}
where $N\cdot v=v_{0}$ for any vector $v$. When this propagator is connected to a fast right moving object, so that $k_{+}^{2}\gg \ul{k}^{2},\, k_{-}^{2}$, the dominant part in \eq{eq:coulomb} is $D_{-\beta}=ik_{\beta}^{\perp}/(k^2k_{+})$ which coincides with the one from the \eq{eq:lightcone} in $A_{+}=0$ gauge except for the $i\epsilon$ pole. Similarly for a left moving system Coulomb gauge is equivalent to $A_{-}=0$ gauge. Therefore, in Coulomb gauge the gluon fields emitted from a right moving system are in $A_{+}=0$ gauge, while for a left moving system in $A_{-}=0$ gauge.

However, the Coulomb gauge does not indicate an explicit choice of $i\epsilon$'s in the propagator. So if we want to choose Coulomb gauge as our overall gauge choice we have to show that LC calculation is independent of the $i\epsilon$ prescriptions in the propagator. Since we will use two different LC propagators for the two colliding nuclei, there are totally four different choices of $i\epsilon$'s  but only three of them are different. Different choices of $i\epsilon$'s correspond to different ways that gluons propagate, i.e. different way of calculating diagrams. In principle we can compare diagrams coming from different $i\epsilon$ choices order by order. Up to tree level diagrams, we manage to show that three different choices of $i\epsilon$'s give an identical result. For the problem at hand we do not need to go beyond tree level calculation, since it is sufficient to compare tree level diagrams in the quasi-classical approach. Part of our goal here is quite similar to what has already been done by the authors in Ref. \cite{Belitsky}, where they introduced a gauge invariant transverse momentum-dependent parton distribution and showed that the definition is independent of the $i\epsilon$ prescriptions on the LC propagator and is the same as the calculation done in covariant gauge. Here, however, we are not only able to show that WW gluon distribution is independent of the $i\epsilon$ prescriptions, but also that the cross section for scalar particle production is independent of the prescriptions and factorized.

\section{Light cone calculation of  scalar particle production}
\label{sec:calculation}
In this section we evaluate scalar particle production in a nucleus-nucleus collision in light cone gauge. In light cone gauge noncausal interactions can happen. In our problem the dominant part of the LC propagator is $D_{\alpha\beta}^{\phantom{\alpha}\perp}=in_{\alpha}k_{\beta}^{\perp}/[(k^{2}+i\epsilon)(k\cdot n+i\epsilon)]$, which gives rise to the noncausal interactions indicated by the $1/(k\cdot n+i\epsilon)$ factor from the second term of the propagator. So when we speak of the choice of $i\epsilon's$ we always refer to the second term of the propagator. Other components of the LC propagator are small. For example, in $A\cdot n=A_{+}=0$ gauge, $D_{-+}(k)=0$ and $D_{--}(k)=2ik_{-}/k^2k_{+}$ is small compared to $D_{\alpha\beta}^{\phantom{\alpha}\perp}(k)$.
\begin{figure}[h]
  \centering
  \subfigure[]{\label{fig:gaugerotation:a}
    \includegraphics[width=6cm]{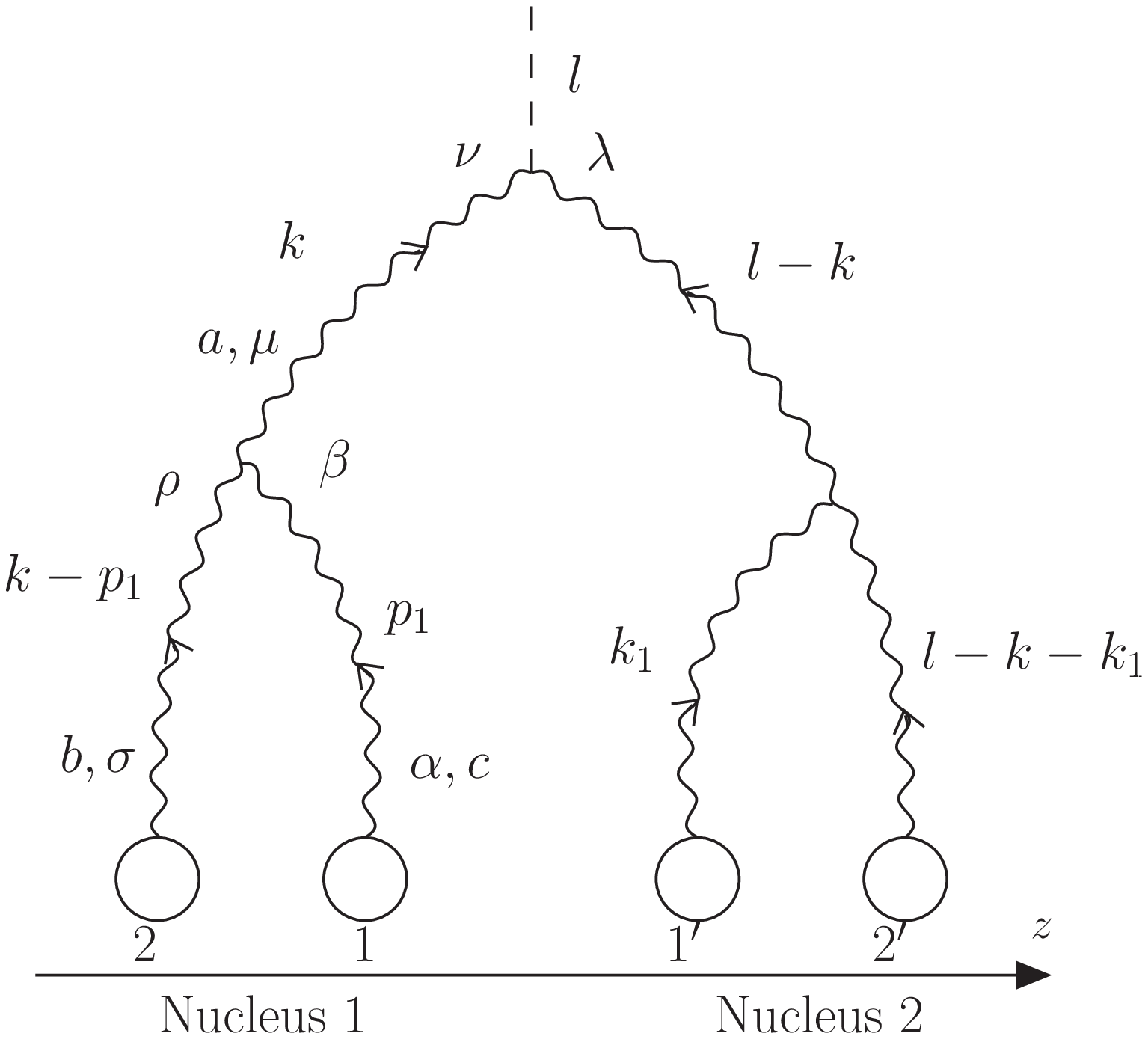}}
  \subfigure[]{\label{fig:gaugerotation:b}
    \includegraphics[width=6cm]{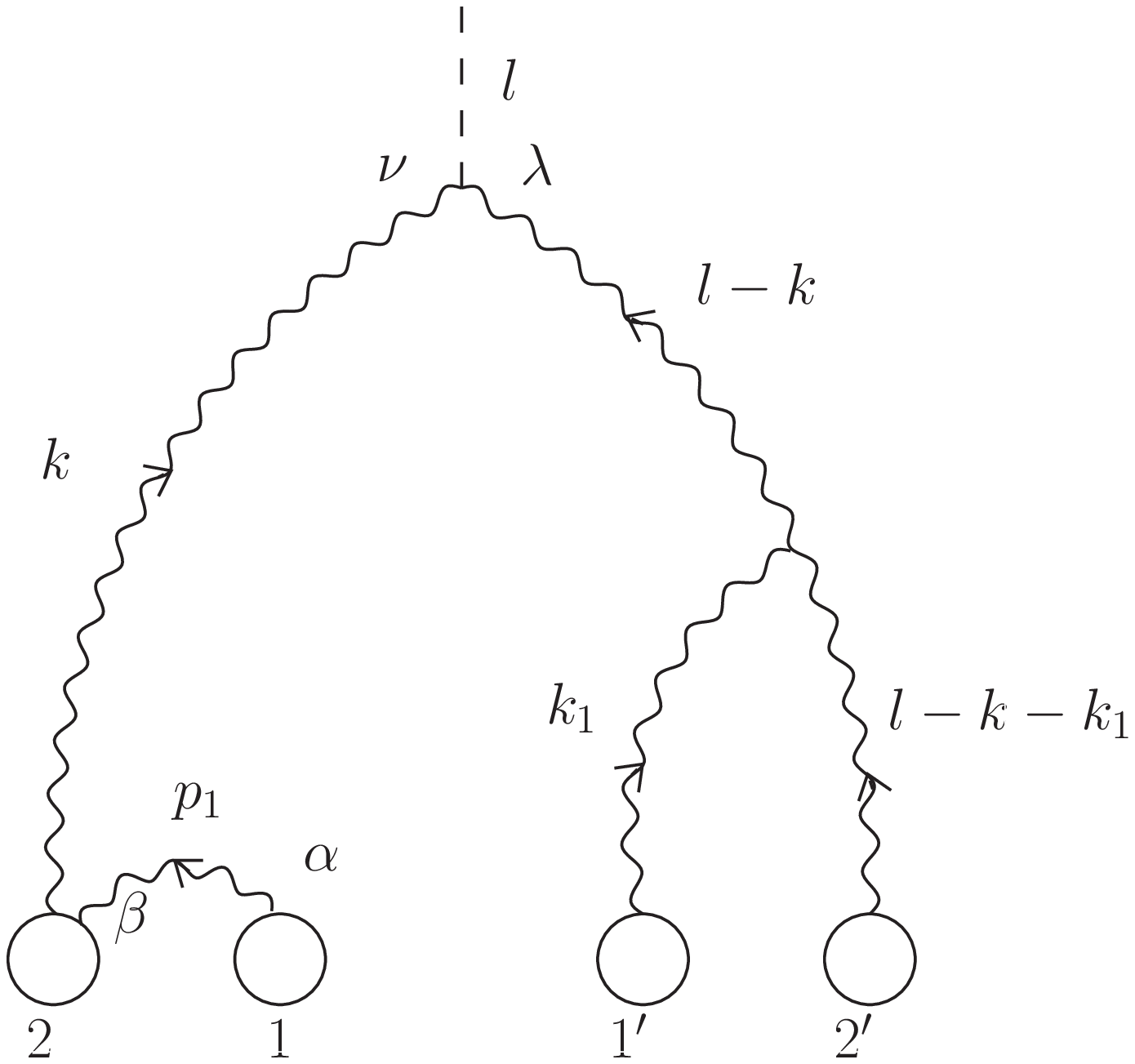}}
  \subfigure[]{\includegraphics[width=6cm]{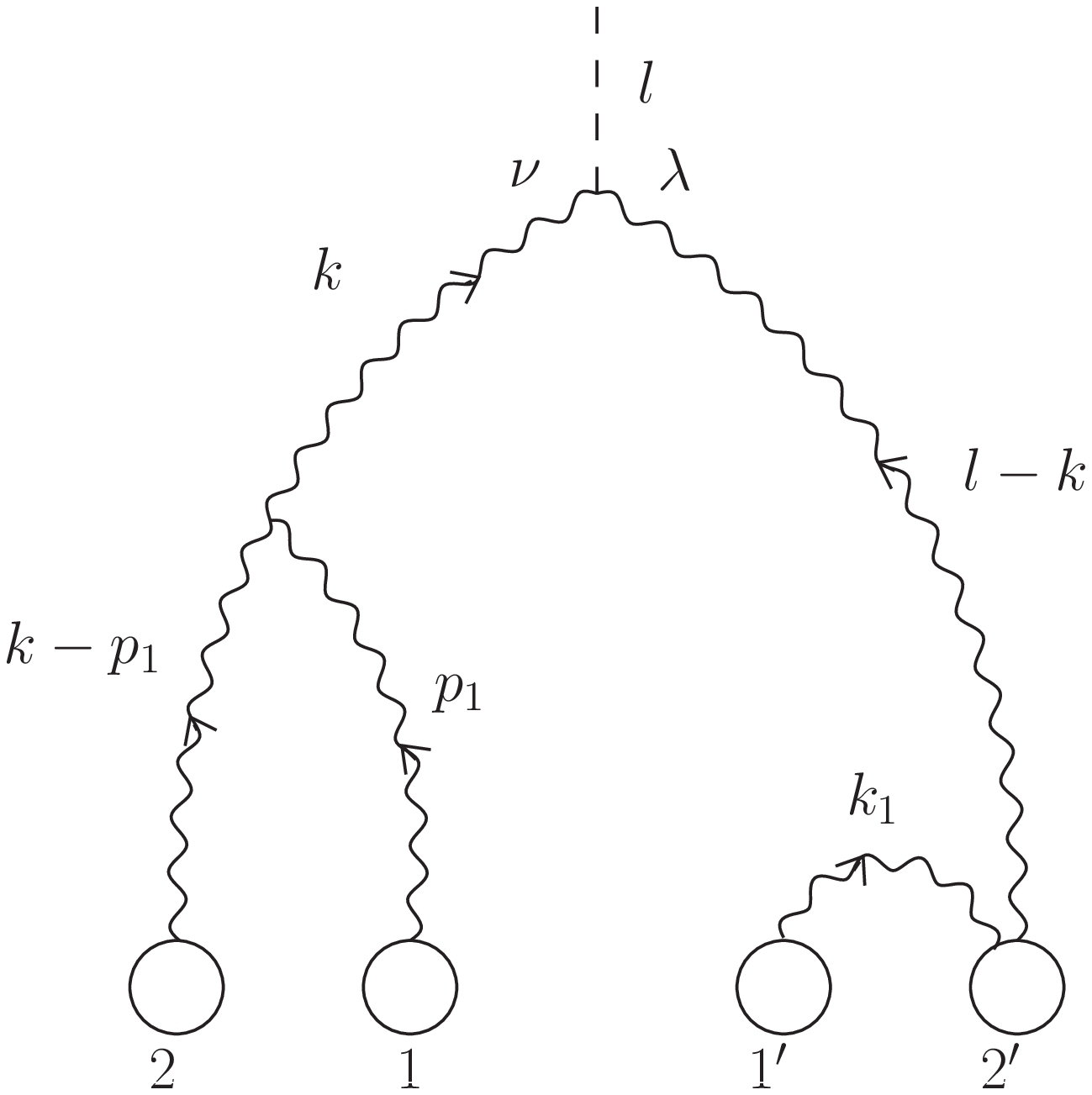}}
  \subfigure[]{\includegraphics[width=5.5cm]{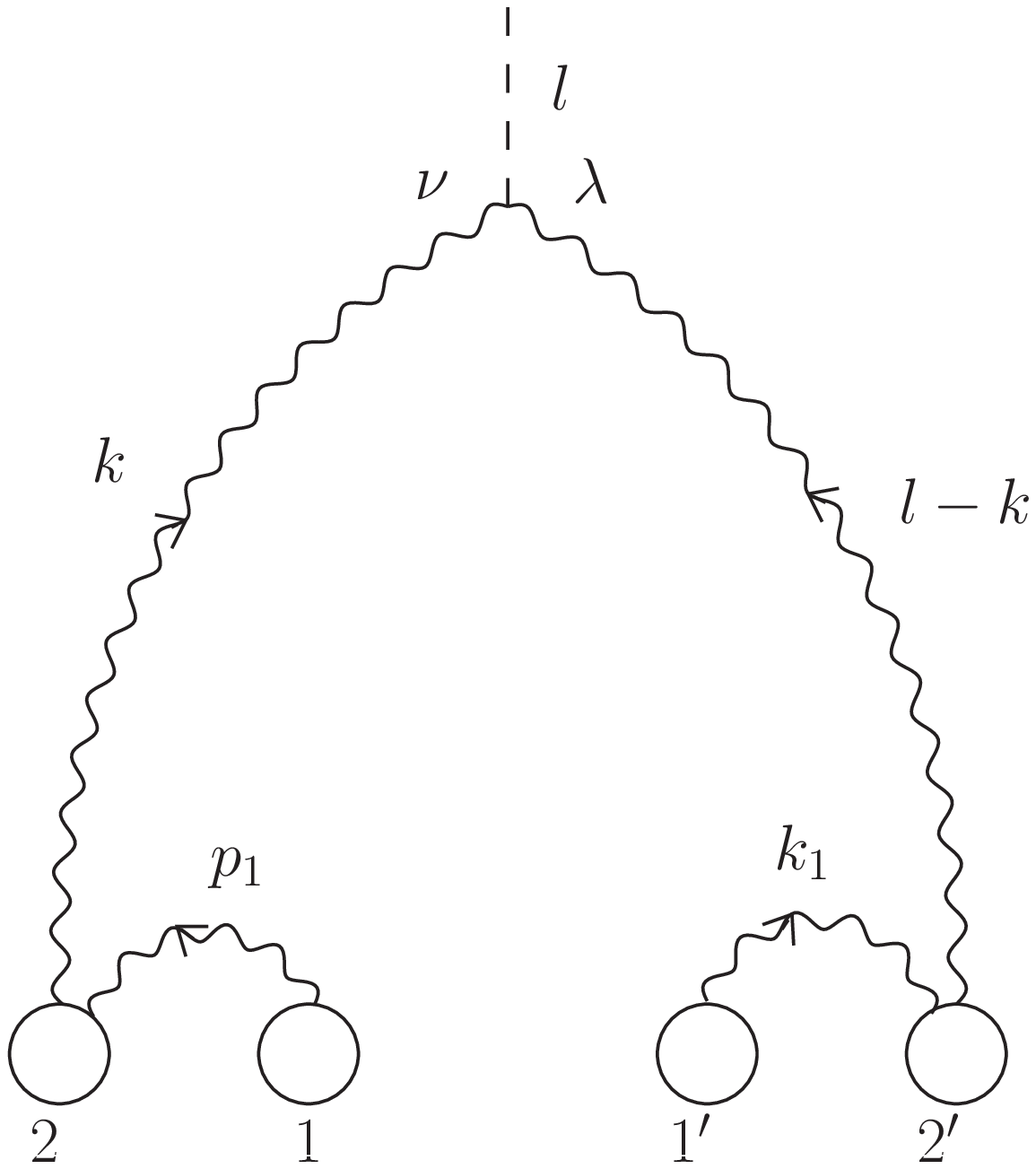}}
\caption{Scalar particle production in light cone gauge involving two nucleons from each nucleus. The hard gluons come from nucleons 2 and $2'$, respectively.}
\label{fig:gaugerotation}
\end{figure}

In order to illustrate the calculation we take two nucleons from each nucleus as shown in \fig{fig:gaugerotation}, where the two nuclei are moving along the $z$-axis, and the dashed line represents the scalar field. As the two nuclei move towards each other at a very high energy, they are highly contracted in the direction of their motions. However, the nucleons inside a nucleus are well-ordered in the $x_{-}$-direction, for the right-moving nucleus, or $x_{+}$-direction, for the left-moving nucleus. We choose the nucleus 1 to be a right-mover which has a big plus momentum component and is in $A_{+}=0$ gauge. We label the front nucleon in the nucleus 1 to be nucleon 1 which has a smaller $x_{-}$-coordinate than that of nucleon 2, $x_{1-}<x_{2-}$. Similarly, nucleus 2 is a left-mover and is in $A_{-}=0$ gauge. The $x_{+}$-coordinate of nucleon $1'$ is less than that of nucleon $2'$, $x'_{1+}<x'_{2+}$. We will use $(k_{+}+i\epsilon)$ and $\ (k_{-}+i\epsilon)$ choices for the propagators in $A_{+}=0$ and $A_{-}=0$ gauge respectively, which means gluon fields propagate from the front to the back in a given nucleus. In other words, gluons in $A_{+}=0$($A_{-}=0$) gauge only propagate from a small $x_{-}$($x_{+}$)-coordinate to a large $x_{-}$($x_{+}$)-coordinate. With this very property a number of diagrams are prohibited at the beginning. A typical group of diagrams is shown in \fig{fig:gaugerotation}, where we only draw one gluon from each nucleon. It is straightforward to add more gluons, but no more than two gluons from each nucleon in the quasi-classical approximation. If one wants to connect gluon $k_{1}$ with, for example in \fig{fig:gaugerotation:a}, gluon $k$, which come from nucleon 2 that has a smaller $x_{-}$-coordinate than that of nucleon $1'$, then one obtains a diagram that violates the $(k_{+}+i\epsilon)$ prescription. Such diagrams correspond to initial state interactions before the collision, they are suppressed by the mass $M$ in the current $i\epsilon$ prescriptions. A more detail discussion will be given in the appendix. Moreover, in the appendix we repeat the calculation in a totally different $i\epsilon$ prescription, which leads to a huge amount of initial state interactions, then such diagrams are necessary in the calculation. In principle the current $i\epsilon$ prescriptions would lead to severe final state interactions, however, color neutral scalar particle knows nothing about the gluon fields after the collision once produced, clearly it is not the case for gluon production. Therefore the four diagrams shown in \fig{fig:gaugerotation} are the only diagrams we have to calculate if we allow only one gluon from each nucleon. Moreover, hard gluons have to come from the last interacting nucleon in each nucleus according to the same reasons. In our approximation, we only need one hard gluon from each nucleus to make up the large mass $M$, and one-hard-gluon approximation is also one of the characteristics of the WW gluon fields.
\begin{figure}[h]
  \centering
  \subfigure[]{  \label{fig:gauge}
  \includegraphics[width=6cm]{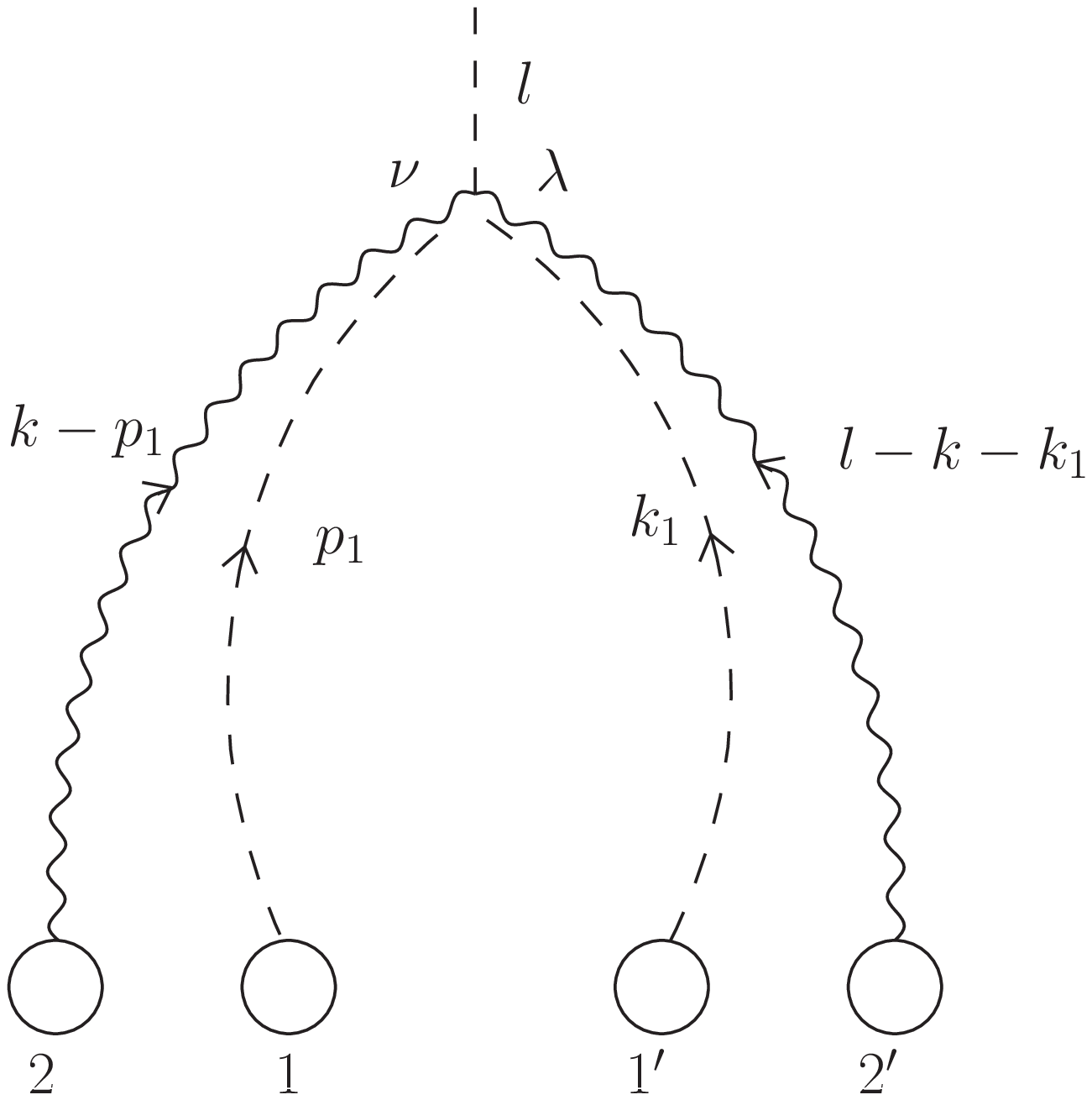}}
  \subfigure[]{   \label{fig:ww}
    \includegraphics[width=6cm]{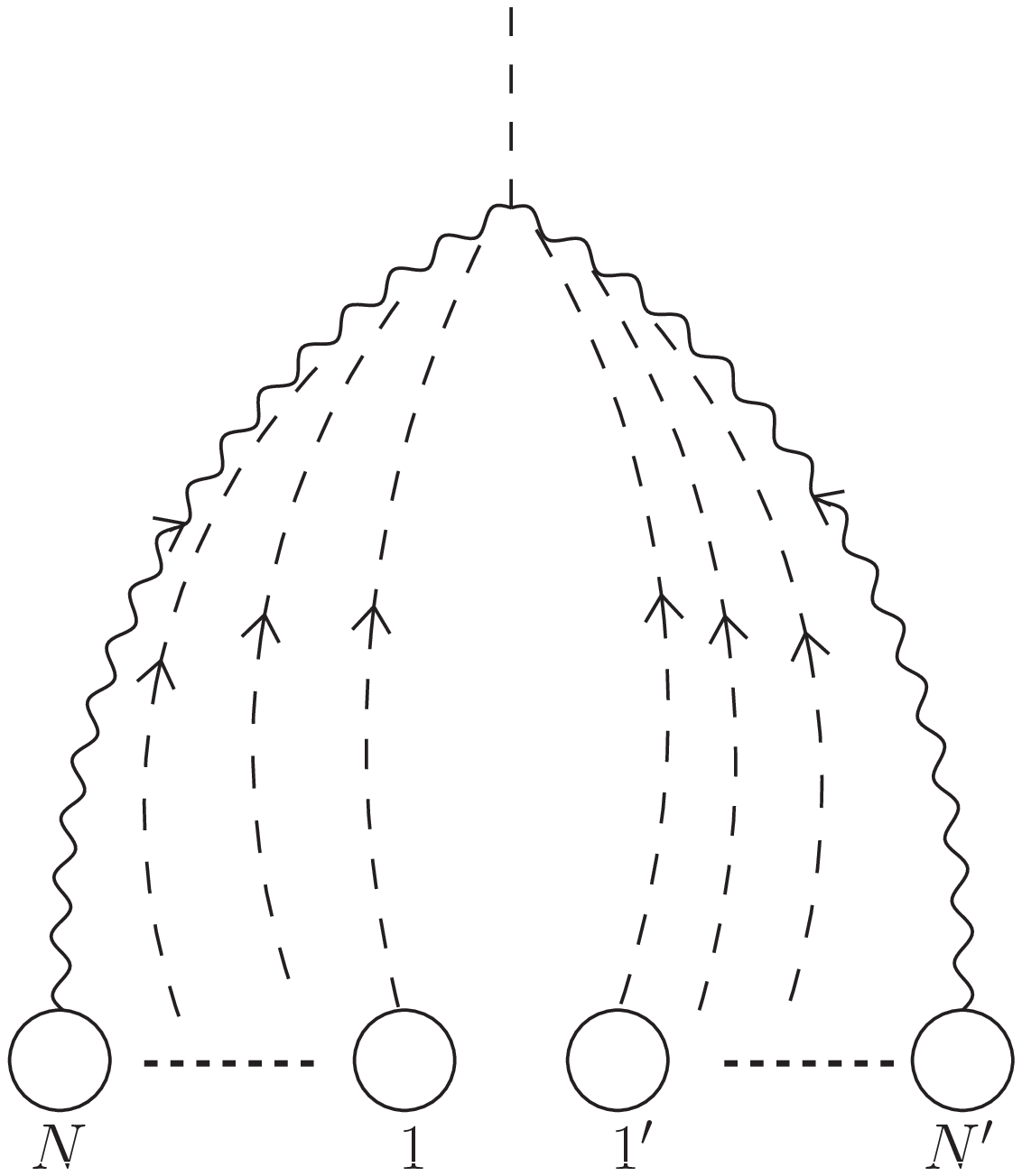}}
  \caption{Scalar particle production in the language of gauge rotation involving two nucleons (a) and an arbitrary number of nucleons (b) in each nucleus. A dashed line with an arrow denotes the gauge rotation. }
\end{figure}

We will use the STW identities to evaluate the graphs in \fig{fig:gaugerotation}. First, we focus on the left part of \fig{fig:gaugerotation:a}. The relevant factors are
\begin{equation}
  \label{eq:example}
  \begin{split}
  \int d p_{1+}e^{ip_{1+}(x_{1-}-x_{2-})}&\frac{i}{p^{2}_{1}+i\epsilon}\frac{\eta_{\alpha} p_{1\beta}^{\perp}}{p_{1+}+i\epsilon}\frac{i}{(k-p_{1})^{2}+i\epsilon}\frac{\eta_{\sigma}(p_{1}-k)_{\rho}^{\perp}}{p_{1+}-k_{+}-i\epsilon}\Gamma_{\rho\beta\mu}\\
  &\times\frac{-i}{k^{2}+i\epsilon}\bigg[g_{\mu\nu}-\frac{\eta_{\mu}k_{\nu}}{k_{+}+i\epsilon}-\frac{\eta_{\nu}k_{\mu}}{k_{+}-i\epsilon}\bigg]v_{\nu\lambda},
\end{split}
\end{equation}
where $\Gamma_{\rho\beta\mu}=gf_{abc}\big[g_{\beta\rho}(2p_{1}-k)_{\mu}+g_{\rho\mu}(2k-p_{1})_{\beta}+g_{\mu\beta}(-k-p_{1})_{\rho}\big]$ is the usual three-gluon vertex, $v_{\nu\lambda}$ is the scalar-gluon vertex and $\eta\cdot k=k_{+}$. The dominant part of the $p_{1}$-propagator is proportional to $p_{\beta}^{\perp}/(p_{1+}+i\epsilon)$, while the $(k-p_{1})$-propagator contains $(p_{1}-k)_{\rho}^{\perp}/(p_{1+}-k_{+}-i\epsilon)$. The $p_{1+}$-poles only come from the two propagators and are on opposite sides of the real axis. We now carry out an integration over $p_{1+}$, the phase factor $e^{ip_{1+}(x_{1-}-x_{2-})}$ with $x_{1-}-x_{2-}<0$ tells us to distort the contour in the lower-half plane and pick up the pole in the $p_{1}$-propagator which sets $p_{1+}=0$. Since $p_{1-}$ is very small compared to $l_{-}$ for scalar particle production in a central rapidity region, we can replace $p_{1\beta}^{\perp}$ by $p_{1\beta}$ which allows us to apply the STW identities to the $\beta$-index. The $p_{1\beta}$ factor will separate the $\Gamma_{\rho\beta\mu}$ vertex into two different parts
\begin{equation}
  \label{eq:pline}
  p_{1\beta}\Gamma_{\rho\beta\mu}=gf_{abc}\Big\{\big(g_{\rho\mu}k^2-k_{\rho}k_{\mu}\big)-\big[g_{\rho\mu}(p_{1}-k)^2-(p_{1}-k)_{\mu}(p_{1}-k)_{\rho}\big]\Big\}.
\end{equation}

The first term in \eq{eq:pline} multiplied by the $k$-propagator gives $g_{\rho\nu}-\eta_{\rho}k_{\nu}/(k_{+}+i\epsilon)$. However, the $\eta_{\rho}$ term is eliminated by the $(k-p_{1})$-propagator; i.e. $(k-p_{1})^{\perp}_{\rho}\eta_{\rho}=0$, so only $g_{\rho\nu}$ is left. This corresponds to eliminating the $k$-propagator and bringing the $p_{1}$-line to the gluon-scalar vertex. Similarly, the second term in \eq{eq:pline} eliminates the $(k-p_{1})$-propagator and brings the $p_{1}$-line back to the nucleon 2. Due to color factors the second term cancels with the graph shown in \fig{fig:gaugerotation:b}. Therefore, the non-vanishing contribution comes from the first term which is denoted by a dashed line with an arrow, a notation introduced by t' Hooft \cite{wardidentities}, attaching to the scalar-gluon vertex. Similar calculations can also be found in Ref. \cite{recombination,gaugerotation,pA}. The new $p_{1}$-line changes the momentum of the $(k-p_{1})$-line to $k$, so there is no $p_{1}$-dependence at the vertex $v_{\nu\lambda}$. This gluon line can be interpreted as a gauge rotation which is essentially a soft gluon, or a small-$x$ gluon which is only coherent in the longitudinal direction. The gauge rotation brings color as well as transverse momentum but no longitudinal momentum to the vertex. We can also apply the same technique to the $k_{1}$-line which also becomes a gauge rotation. The sum of the four diagrams in \fig{fig:gaugerotation} can be greatly simplified by the STW identities and become the one diagram shown in \fig{fig:gauge} with all the soft gluons interpreted as gauge rotations. Now the vertex part in \fig{fig:gauge} reads
\begin{equation}
  \begin{split}
  (k-p_{1})_{\nu}^{\perp}(l-k-k_{1})_{\lambda}^{\perp}v_{\nu\lambda}&\propto  (k-p_{1})_{\nu}^{\perp}(l-k-k_{1})_{\lambda}^{\perp}\big[g_{\nu\lambda}k\cdot (l-k)-k_{\lambda}(l-k)_{\nu}\big]\\
  &\approx -(\ul{k}-\ul{p}_{1})\cdot(\ul{l}-\ul{k}-\ul{k}_{1})l_{+}l_{-},
\end{split}
\end{equation}
where the transverse momentum components come from the two hard gluon propagators. We always assume that the mass squared of the scalar particle $M^{2}=l^2$ is larger than any transverse momentum in the system. We see that although there are two soft gluons attached to the hard gluons, only the transverse momenta of the hard lines enter the vertex, which is the very characteristic of the WW field $A_{\mu}^{WW}$ in momentum space. Therefore, we can identify the hard gluons from each nucleus as the WW fields at this order. This identification become even sharper by noticing that the gluon field emerging from one nucleus, either the left or right part of \fig{fig:gauge}, is the same as the result of Fig. 4 in Ref. \cite{gaugerotation}.
\begin{figure}[h]
  \centering
  \includegraphics[width=8cm]{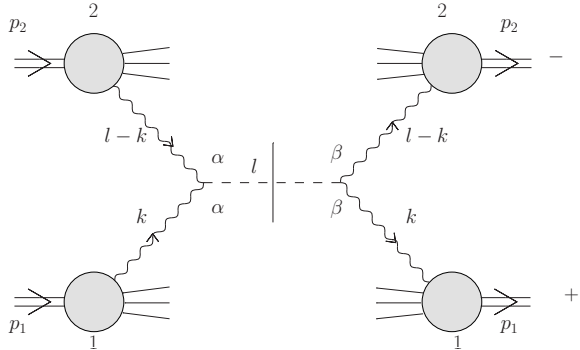}
  \caption{Scalar particle production in nucleus-nucleus collisions. The gluon fields from each nucleus are interpreted as the WW fields.}
  \label{fig:nn}
\end{figure}

This low order calculation can be easily generalized to an arbitrary number of interacting nucleons in each nucleus. The hard gluons still come from the last interacting nucleons while all the soft gluons are propagating backwards. We always apply the STW identities to the soft gluons that come from the most front nucleons first. The soft gluons become gauge rotations and merge at the scalar-gluon vertex as shown in \fig{fig:ww}. Adding more nucleons corresponds to attaching more gauge rotations to the hard gluons. All the gauge rotations can be resummed in coordinate space, then the hard gluons become the WW fields $A^{WW}_{\mu}(\ul{x})$ of the nuclei \cite{gaugerotation}. Therefore, one can first calculate the scattering amplitude at the lowest order where there is only one gluon from each nucleus as shown in \fig{fig:nn}. This gives
\begin{equation}
  \mathcal{M}=-g^2(4p_{1+}p_{2-})T^{a}_{1}T^{a}_{2}\frac{(\ul{l}-\ul{k})\cdot \ul{k}}{(\ul{l}-\ul{k})^2\ul{k}^2}.
  \label{eq:amplitude}
\end{equation}

We identify the gluon fields in \eq{eq:amplitude} as $A_{\mu}^{\perp}(k)=gT^{a}k_{\mu}^{\perp}/(\ul{k}^2k_{\pm})$ in $A_{\pm}=0$ gauge\cite{overview}, respectively. The cross section is given by
\begin{equation}
  \frac{d \sigma}{d^2\ul{l}dy}=\frac{g_{\phi}^{2}}{4}k_{+}^{2}l_{-}^{2}\int \frac{d^2\ul{k}}{(2\pi)^5}\mathrm{Tr}\big[A_{1\mu}^{\perp}(k)A_{1\nu}^{\perp}(-k)\big]\mathrm{Tr}\big[A_{2\mu}^{\perp}(l-k)A_{2\nu}^{\perp}(-l+k)\big].
\end{equation}
Finally, going to coordinate space 
\begin{equation}
  \begin{split}
    A_{1\mu}^{\perp}(\ul{k},k_{+})=\int d^{2}\ul{x}_{1}\int d x_{1-}\, e^{i(k_{+}+i\epsilon)x_{1-}-i\ul{k}\cdot \ul{x}_{1}}A_{1\mu}^{\perp}(\ul{x}_{1},x_{1-}),\\
    A_{2\mu}^{\perp}(\ul{k},k_{-})=\int d^{2}\ul{x}_{2}\int d x_{2-}\, e^{i(k_{-}+i\epsilon)x_{2+}-i\ul{k}\cdot \ul{x}_{2}}A_{2\mu}^{\perp}(\ul{x}_{2},x_{2+}),
  \end{split}
\end{equation}
one can replace $A_{\mu}^{\perp}$ by the full WW $A_{\mu}^{WW}$ field which resums all the soft gluons. This gives
\begin{equation}
  \frac{d^{3} \sigma}{d^2 \ul{l}dy}=\frac{\pi^3g_{\phi}^2}{8}\int d^{2}\ul{x}e^{i\ul{l}\cdot\ul{x}}\,\tilde{N}_{1\alpha\beta}(\ul{x})\tilde{N}_{2\alpha\beta}(\ul{x})\qquad (\alpha,\beta=1,2),
  \label{eq:csfull}
\end{equation}
which is a factorized form involving two unintegrated gluon distributions
\begin{equation}
  \tilde{N}_{i\alpha\beta}(\ul{x})=-\frac{1}{2\pi^3}\int d^2\ul{b}_{i}\,\mathrm{Tr}[A_{i\alpha}^{WW}(\ul{b_{i}})A_{i\beta}^{WW}(\ul{b_{i}}+\ul{x})]\qquad (i=1,2),
\label{eq:unint}
\end{equation}
where the index $i$ labels the different nucleus. 

We will evaluate \eq{eq:unint} following the procedure outlined in \cite{pA}. In $A_{+}=0$ gauge $A_{\alpha}^{WW}(\ul{x})$ can be written as
\begin{equation}
  A^{WW}_{\alpha}(\ul{x},x_{-})=\int S(\ul{x},b_{-})T^{a}S^{-1}(\ul{x},b_{-})\frac{(\ul{x}-\ul{b})_{\alpha}}{|\ul{x}-\ul{b}|^{2}}\hat{\rho}^{a}(\ul{b},b_{-})\theta(b_{-}-x_{-})d^2\ul{b}db_{-},
\label{eq:afield}
\end{equation}
with
\begin{equation}
  S(\ul{x},x_{-})=P\exp\bigg\{igT^{a}\int\ln \big[|\ul{x}-\ul{b}|\mu\big]\hat{\rho}^{a}(\ul{b},b_{-})\theta(b_{-}-x_{-})d^{2}\ul{b}db_{-}\bigg\},
  \label{eq:s}
\end{equation}
as taken from \cite{pA}. Put $A_{\alpha}^{WW}$ in \eq{eq:unint}, and use
\begin{equation}
  \langle\hat{\rho}^{a}(\ul{b},b_{-})\hat{\rho}^{a'}(\ul{b'},b'_{-})\rangle=\frac{\rho(\ul{b},b_{-})}{N^{2}_{c}-1}\delta(\ul{b}-\ul{b}')\delta(b_{-}-b'_{-})\delta_{aa'}Q^{2}\frac{\partial}{\partial Q^2}xG(x,Q^2),
\end{equation}
to average the color charges. In the McLerran-Venugopalan model we do not have either $Q^2$-dependence or $x$-dependence in $Q^2(\partial/\partial Q^2)xG(x,Q^2)$. One finds
\begin{equation}
  \label{eq:N}
  \begin{split}
    \tilde{N}_{\alpha\beta}(\ul{x})&=-\frac{1}{2\pi^3}\int d^2\ul{b} \int d^2 \ul{b}'db'_{-}\frac{(\ul{b}-\ul{b}')_{\alpha}}{|\ul{b}-\ul{b}'|^2}\frac{(\ul{b}+\ul{x}-\ul{b}')_{\beta}}{|\ul{b}+\ul{x}-\ul{b}'|^2}\frac{\rho(\ul{b}',b'_{-})}{N^{2}_{c}-1}Q^2\frac{\partial}{\partial Q^2}xG(x,Q^2)\\
    &\phantom{==}\times \langle \mathrm{Tr}\big[S(\ul{b},b'_{-})T^{a}S^{-1}(\ul{b},b'_{-})S(\ul{b}+\ul{x},b'_{-})T^aS^{-1}(\ul{b}+\ul{x},b'_{-})\big]\rangle.
  \end{split}
\end{equation}

$\rho(\ul{b}',b'_{-})=\rho_{\mathrm{rel}}$ is the normal nuclear density in the boosted frame and in light cone variables. One can do the $d^{2}\ul{b}'$ integration as
\begin{equation}
  \begin{split}
    \int d^{2}\ul{b}'&\frac{(\ul{b}-\ul{b}')_{\alpha}}{|\ul{b}-\ul{b}'|^2}\frac{(\ul{b}+\ul{x}-\ul{b}')_{\beta}}{|\ul{b}+\ul{x}-\ul{b}'|^2}=\int d^{2}\ul{k}\, e^{-i\ul{k}\cdot \ul{x}}\frac{k_{\alpha}k_{\beta}}{(\ul{k}^2+\mu^2)^{2}}\\
    &=-\pi\bigg[-\delta_{\alpha\beta}K_{0}(|\ul{x}|\mu)+\mu\frac{\ul{x}_{\alpha}\ul{x}_{\beta}}{|\ul{x}|}K_{1}(|\ul{x}|\mu)\bigg]\xrightarrow{\mu\rightarrow 0}-\pi\big[\delta_{\alpha\beta}\ln (|\ul{x}|\mu)+\hat{\ul{x}}_{\alpha}\hat{\ul{x}}_\beta\big].
  \end{split}
\end{equation}
 $\mu$ is some infrared cut-off and $\hat{\ul{x}}_{\alpha}=\ul{x}_{\alpha}/|\ul{x}|$ is a unit vector. The trace is evaluated by expanding nucleons from $S(\ul{x},x_{-})$ one by one, the result is also taken from \cite{pA}
 \begin{equation}
   \begin{split}
   \langle \mathrm{Tr}&\big[S(\ul{b},b'_{-})T^{a}S^{-1}(\ul{b},b'_{-})S(\ul{b}+\ul{x},b'_{-})T^aS^{-1}(\ul{b}+\ul{x},b'_{-})\big]\rangle\\
   &=C_{F}N_{c}\exp\bigg[-g^{2}\frac{\pi\rho_{\mathrm{rel}}N_{c}\ul{x}^{2}}{4(N^{2}_{c}-1)}xG(x,1/\ul{x}^{2})(b'_{-}+b'_{0-})\bigg],
 \end{split}
\end{equation}
where $\pm b'_{0-}$ is the upper (lower) limit of the $b_{-}$ integration in \eq{eq:s}. Plugging this back in \eq{eq:N} and performing the $db^{'}_{-}$ integration and using the nuclear density in the center of mass frame $\rho=\rho_{\mathrm{rel}}/\gamma\sqrt{2}$ with $\gamma$ the Lorentz contraction factor, one obtains
\begin{equation}
  \tilde{N}_{i\alpha\beta}(\ul{x})=\frac{N_{c}^{2}-1}{4\pi^4\alpha N_{c}\ul{x}^{2}}\frac{1}{2}\bigg[\delta_{\alpha\beta}+\frac{\hat{x}_{\alpha}\hat{x}_{\beta}}{\ln(|\ul{x}|\mu)}\bigg]\int d^{2}\ul{b}_{i}\,(1-e^{-Q_{is}^2\ul{x}^2/4}),
\label{eq:n}
\end{equation}
where $Q_{is}^2$ are the corresponding saturation momenta for the two nuclei. For a spherical nucleus of radius $R$, it can be written as $Q_{s}^2=8\pi^2 \alpha N_{c}\sqrt{R^2-\ul{b}^2}\rho x G(x,1/\ul{x}^{2})/(N^{2}_{c}-1)$.

We can go to momentum space by
\begin{equation}
  N_{i\alpha\beta}(\ul{k}_{i})=\int d^2\ul{x}\, e^{i\ul{k}_{i}\cdot \ul{x}}\tilde{N}_{i\alpha\beta}(\ul{x}).
\end{equation}
The diagonal part of the above expression are the usual WW gluon distributions in momentum space, i.e. $N_{i}(\ul{k}_{i})=\delta_{\alpha\beta}N_{i\alpha\beta}(\ul{k}_{i})$. The off-diagonal part are different gluon distributions, the so-called linearly polarized gluon distribution, also found in \cite{linearlypolarized}
\begin{equation}
  n_{i}(\ul{k}_{i})=(2\hat{\ul{k}}_{i\alpha}\hat{\ul{k}}_{i\beta}-\delta_{\alpha\beta})N_{i\alpha\beta}(\ul{k}_{i})=\frac{N_{c}^2-1}{4\pi^3 \alpha N_{c}}\int d|\ul{x}|d^2\ul{b}_{i}\,\frac{J_{2}(|\ul{k}_{i}||\ul{x}|)}{|\ul{x}|\ln(1/|\ul{x}|\mu)}\big(1-e^{-Q^{2}_{is}\ul{x}^2/4}\big).
\label{eq:linear}
\end{equation}
$\hat{\ul{k}}_{i\alpha}$ are unit vectors in momentum space and $J_{n}(x)$ is the Bessel function of the first kind. We see that $N_{\alpha\beta}(\ul{k})$ can be separated into two different unintegrated distribution functions, i.e. $N_{i\alpha\beta}(\ul{k}_{i})=\frac{1}{2}\delta_{\alpha\beta}N_{i}(\ul{k}_{i})+(\hat{\ul{k}}_{i\alpha}\hat{\ul{k}}_{i\beta}-\frac{1}{2}\delta_{\alpha\beta})n_{i}(\ul{k}_{i})$. This decomposition is usually encountered in studies of transverse momentum dependent factorization, see for example \cite{Qiu,Boer}. We can rewrite the cross section in terms of these two gluon distributions in momentum space as
\begin{equation}
  \label{eq:fullcs}
  \frac{d^3\sigma}{d^2\ul{l}dy}=\frac{g_{\phi}^{2}\pi}{64}\int d^2\ul{k}_{1}d^2\ul{k}_{2}\, \delta^{(2)}(\ul{l}-\ul{k}_{1}-\ul{k}_{2})\Big\{N_{1}(\ul{k}_{1})N_{2}(\ul{k}_{2})+\big[2(\hat{\ul{k}}_{1}\cdot \hat{\ul{k}}_{2})-1\big]n_{1}(\ul{k}_{1})n_{2}(\ul{k}_{2})\Big\}.
\end{equation}
A simple version of \eq{eq:fullcs} is found earlier by authors \cite{higgs}, where they calculate Higgs production in pA collisions. 

In the leading logarithmic approximation and when $\ul{l}^2\sim Q_{s}^2$, gluons with $\ul{k}^{2}_{i} \lesssim Q_{s}^{2}$ give the dominant contribution to the cross section and the factor $\ln(|\ul{x}|\mu)$ is large. We only need to keep the $\delta_{\alpha\beta}$ term in \eq{eq:n}, then obtain our main result
\begin{equation}
  \label{eq:cs}
  \frac{d^3 \sigma}{d^2\ul{l}dy}=\frac{g_{\phi}^2\pi^3}{16}\int d^2\ul{x}e^{i\ul{l}\cdot \ul{x}}\,\tilde{N}_{1}(\ul{x})\tilde{N_{2}}(\ul{x}),
\end{equation}
where 
\begin{equation}
  \tilde{N}_{i}(\ul{x})=-\frac{1}{2\pi^3}\int d^2\ul{b}_{i}\, \mathrm{Tr}[A^{WW}_{i}(\ul{b_{i}})\cdot A^{WW}_{i}(\ul{b}_{i}+\ul{x})]=\frac{N_{c}^{2}-1}{4\pi^4\alpha N_{c}\ul{x}^{2}}\int d^{2}\ul{b}_{i}\,\big[1-e^{-Q_{iS}^2\ul{x}^2/4}\big],
\label{eq:unintww}
\end{equation}
are the unintegrated WW gluon distributions of the nuclei. It is not a surprise that WW gluon distribution manifest itself easily in this process, after all the WW gluon distribution counts the number of gluons in the nucleus and is the intrinsic gluon distribution of the nucleus.

When $\ul{l}^2\gg Q_{s}^2$, since the two nuclei contribute equally to the transverse momentum of the scalar particle, we also have $\ul{k}_{i}^2\gg Q_{s}^{2}$. In this limit, $|\ul{x}|$ is very small, the exponential in \eq{eq:unintww} can be expanded and the WW gluon distribution becomes additive
\begin{equation}
  N_{i}(\ul{k})\approx A_{i}\frac{\alpha C_{F}}{\pi^2}\frac{N_{c}}{\ul{k}^2}=A_{i}N_{0}(\ul{k}) \qquad (\ul{k}^2\gg Q_{s}^2),
     \label{eq:largek}
\end{equation}
where 
\begin{equation}
  A_{i}=2\int d^2\ul{b}_{i}\, \rho_{i} \sqrt{R^2-\ul{b}_{i}^{2}}
\end{equation}
are the atomic numbers of the corresponding nucleus and 
\begin{equation}
   N_{0}(\ul{k})=\frac{1}{\pi}\frac{\partial}{\partial \ul{k}^2}\big[xG(x,k^2)\big],
\end{equation}
is the unintegrated gluon distribution of a nucleon at the lowest order. The other gluon distribution has the same behavior,
\begin{equation}
  n_{i}(k)\approx A_{i}\frac{\alpha C_{F}}{\pi^2}\frac{N_{c}}{\ul{k}^2} \qquad (\ul{k}^2\gg Q_{s}^{2}),
\end{equation}
where we have used $\int^{\infty}_{0}d|\ul{x}|\, |\ul{x}|J_{2}(|\ul{k}||\ul{x}|)=2/\ul{k}^2$. Therefore, in this limit we identify the two gluon distributions, then \eq{eq:fullcs} becomes
\begin{equation}
   \frac{d^3\sigma}{d^2\ul{l}dy}=\frac{g_{\phi}^{2}\pi}{64}\int d^2\ul{k}_{1}d^2\ul{k}_{2}\, \delta^{(2)}(\ul{l}-\ul{k}_{1}-\ul{k}_{2})2(\hat{\ul{k}}_{1}\cdot \hat{\ul{k}}_{2})N_{1}(\ul{k}_{1})N_{2}(\ul{k}_{2}),
   \label{eq:csspin}
 \end{equation}
where now $N_{i}(\ul{k})$ take the form indicated in \eq{eq:largek}. Similar expressions have also been obtained in a $k_{t}$-factorization formalism for pp \cite{Lipatov,Hautmann} and in the transverse-momentum-dependent factorization approach for pp \cite{pp} and pA \cite{higgs}.

\section{Covariant gauge calculation}
\label{sec:covariant}
\begin{figure}[h]
  \centering
 \subfigure[]{\label{fig:cgone}
    \includegraphics[width=7cm]{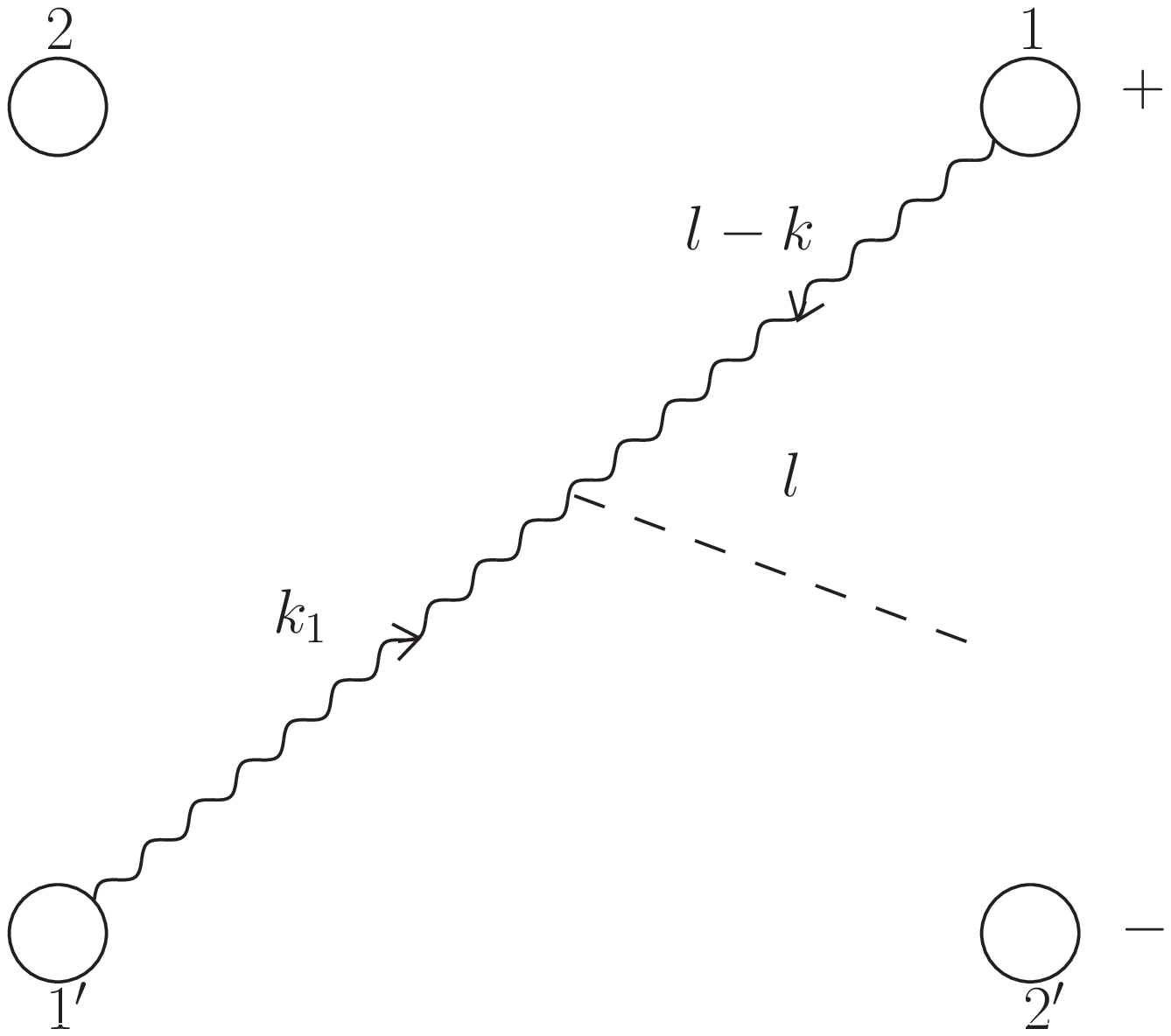}}
 \subfigure[]{\label{fig:cgtwo}
    \includegraphics[width=7cm]{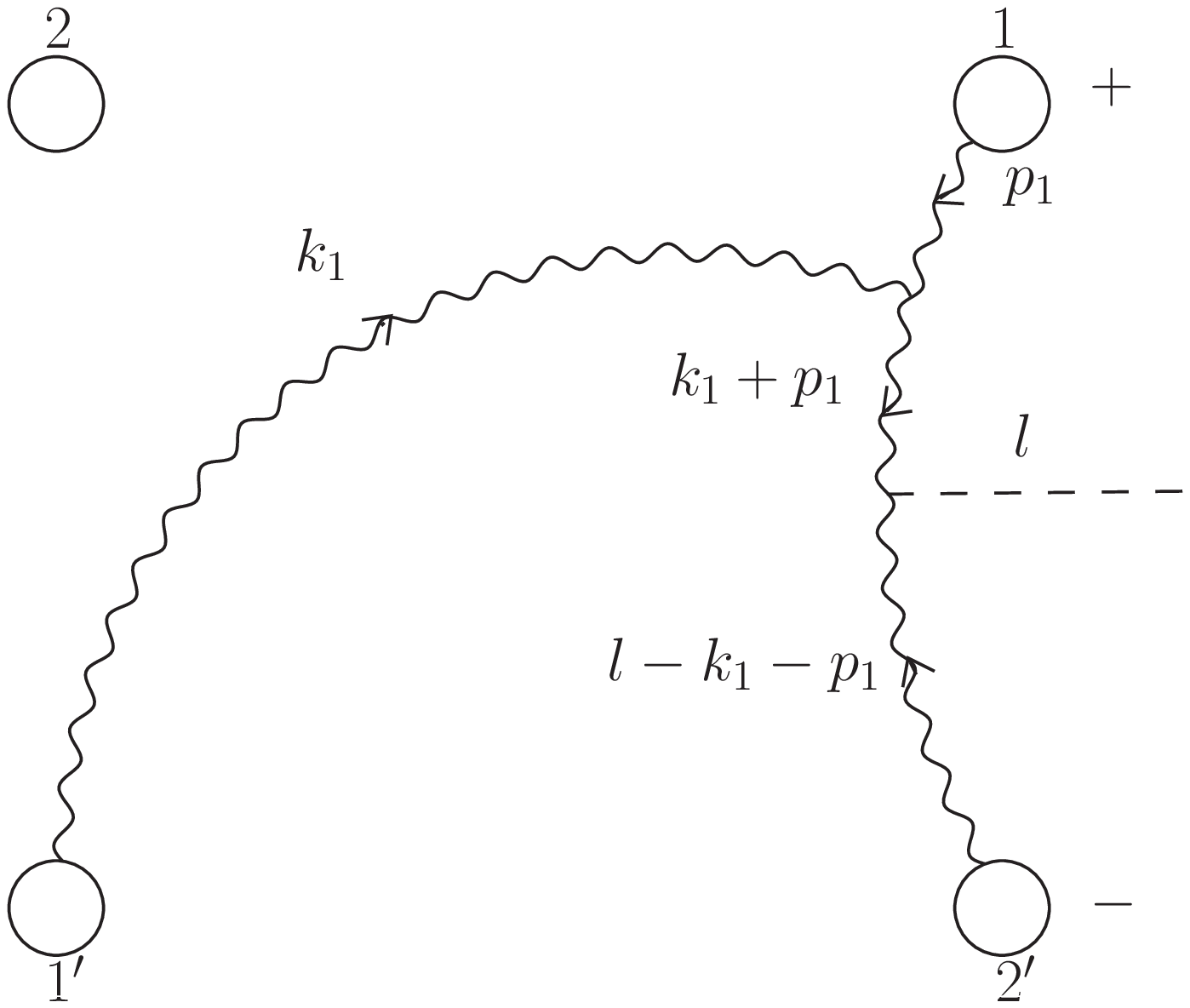}}
 \subfigure[]{\label{fig:cgthree}
    \includegraphics[width=7cm]{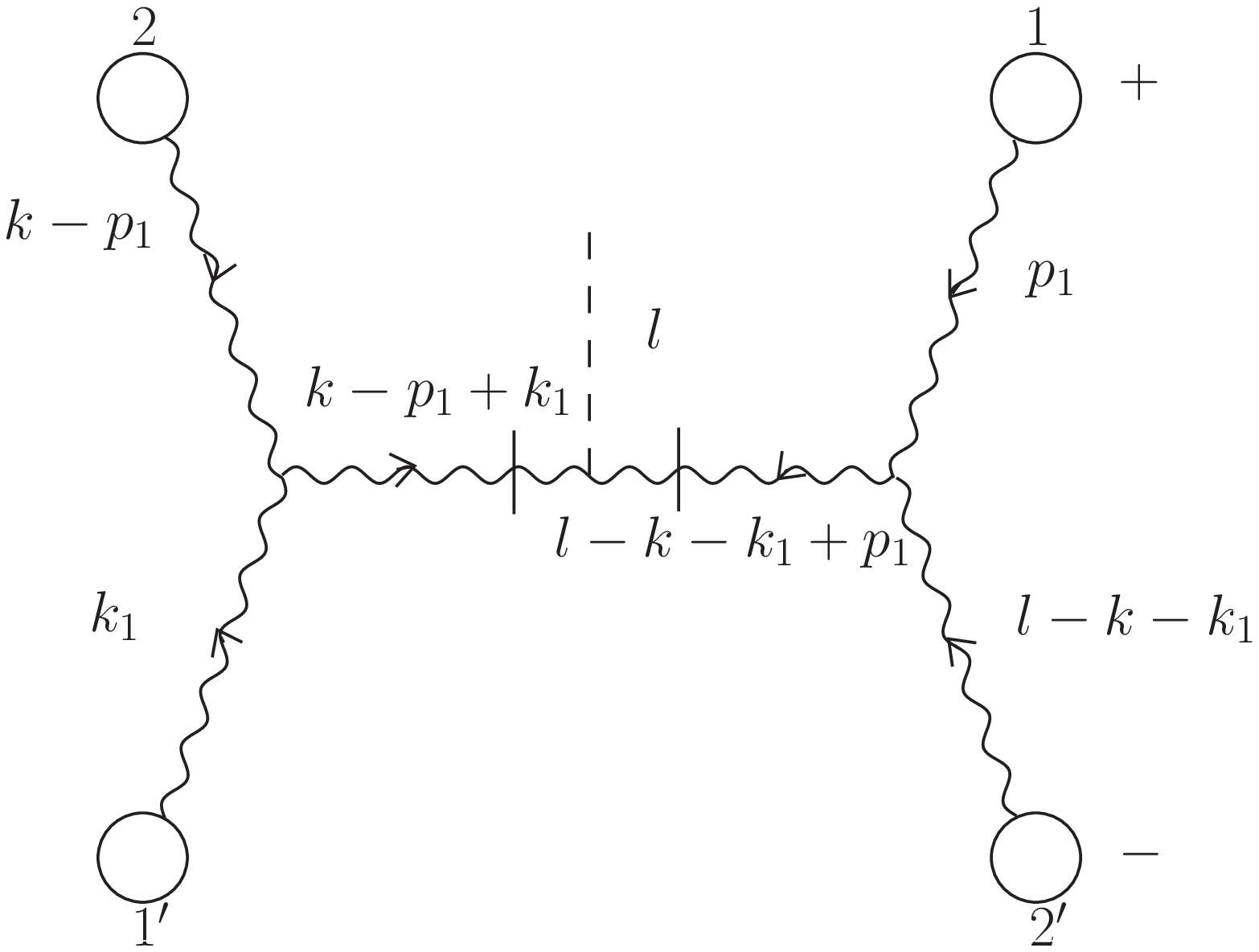}}
 \subfigure[]{\label{fig:cgzero}
    \includegraphics[width=7cm]{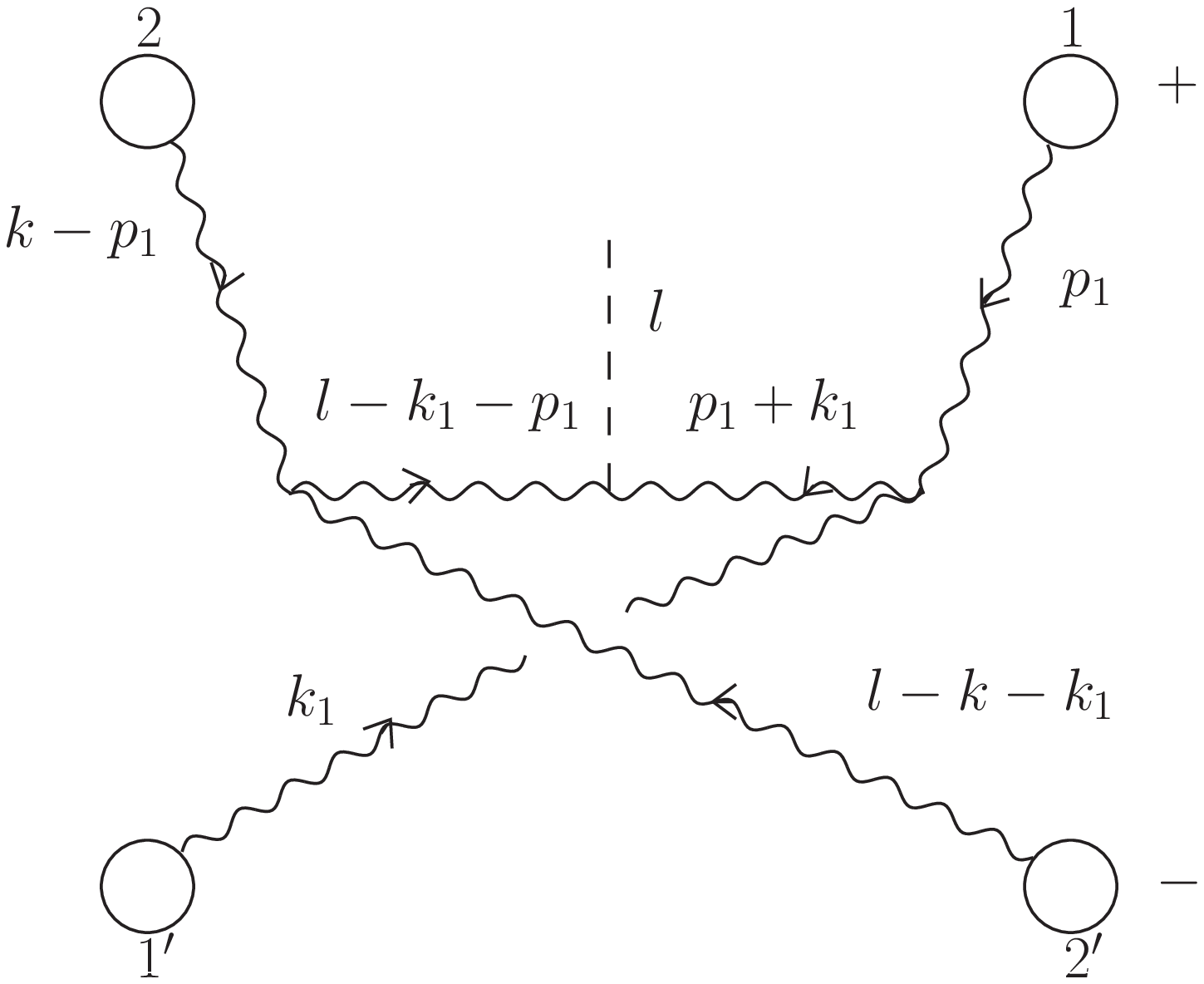}}
  \caption{Scalar particle production in covariant gauge.}
  \label{fig:cg}
\end{figure}

Since we have used an unusual gauge choice to calculate the process, it is interesting to see what the calculation looks like in covariant gauge. As an example take two nucleons from each nucleus and use the same labeling as before. In covariant gauge interactions happen causally and locally. As the two nuclei approach each other, nucleon 1 meets nucleon $1'$ first. One way of producing the scalar particle is to have gluons from them creating the scalar particle, while nucleons 2 and $2'$ act as spectators as shown in \fig{fig:cgone}. The hard gluons can also come from nucleons 1 and $2'$ as shown in \fig{fig:cgtwo}. In this situation the gluon from nucleon 1 can be scattered by nucleon $1'$ before it meets the gluon from nucleon $2'$. A similar process where the hard gluons come from nucleons 2 and $1'$ is also possible. It is not shown in \fig{fig:cg}. Moreover, hard gluons can also be emitted from nucleons 2 and $2'$ as shown in \fig{fig:cgthree}. As the two nuclei pass through each other the gluon from nucleon 2 has to pass nucleon $1'$ first, and one (or two) gluon exchange can take place between them, that is the $k_{1}$-line in \fig{fig:cgthree}. A similar process can happen to the other hard gluon, i.e. the $p_{1}$-line. Finally, those two hard gluons collide and produce the scalar particle. This scattering sequence is equivalent to the diagram in \fig{fig:gauge} in light cone gauge. So, in covariant gauge, a hard gluon from a nucleon will be multiply scattered by nucleons from the other nucleus and receives large transverse momentum from the multiple scattering. In order to realize this picture calculationally it is important to include the phase factors which indicate which nucleons are in front and which are in back \cite{multiplescattering}. The factor is $e^{ip_{1+}(x_{1-}-x_{2-})}e^{ik_{1-}(x'_{1+}-x'_{2+})}$ which dictates how we should distort the contours to pick up the poles in the propagators.

Let us examine in detail the propagators of \fig{fig:cgthree}. In order to generate the large mass $M$ of the scalar particle the large plus momentum, $l_{+}$, comes from nucleon 2 and the large minus momentum,$l_{-}$, comes from nucleon $2'$. Momenta $k_{1}$ and $p_{1}$ are considered to be soft, so we may assume $k_{-},|\ul{k}|\ll l_{-}$ and $p_{1+},|\ul{p}_{1}|\ll l_{+}$. Moreover, we also have $(k-p_{1})_{-}\approx 0$ and $(l-k-k_{1})_{+}\approx 0$ because minus(plus) momentum component of a gluon emitted from a right(left) moving nucleon is very small. Since $k_{1+}\approx 0$, we have $k_{+}\approx l_{+}$. With such a momentum choice the picture of multiple scattering is fulfilled. The relevant integral is 
\begin{equation}
  \label{eq:covcal}
  \begin{split}
  &\int dp_{1+}dk_{1-}\frac{e^{ip_{1+}(x_{1-}-x_{2-})}e^{ik_{1-}(x'_{1+}-x'_{2+})}}{\big[(l-k-k_{1}+p_{1})^2+i\epsilon\big]\big[(k-p_{1}+k_{1})^2+i\epsilon\big]}\\
  &=\frac{1}{4l_{+}l_{-}}\int dp_{1+}\frac{e^{ip_{1+}(x_{1-}-x_{2-})}}{p_{1+}-(\ul{l}-\ul{k}-\ul{k}_{1}+\ul{p}_{1})^2/2l_{-}+i\epsilon}\int dk_{1-}\frac{e^{ik_{1-}(x'_{1+}-x'_{2+})}}{k_{1-}-(\ul{k}-\ul{p}_{1}+\ul{k}_{1})^2/2l_{+}+i\epsilon},
\end{split}
\end{equation}
where we have included two phase factors with $x_{1-}-x_{2-}<0$ and $x'_{1+}-x'_{2+}<0$. With the same approximation other gluon propagators do not have additional poles contributing to \eq{eq:covcal}, for example, we may take $k^2_{1}+i\epsilon\approx -\ul{k}^2_{1}+i\epsilon$ in the $k_{1}$-propagator. The phase factors tell us to do both contour integrations $\int dk_{1-}$ and $\int d p_{1+}$ in the lower half plane to pick up the poles in the corresponding propagators. After the contour integrations, the two gluon propagators are put on shell, indicated by the vertical lines in \fig{fig:cgthree}, then the two successive scatterings are independent. The $k_{1}$- and $p_{1}$-lines become soft and only carry transverse momenta and colors to the hard gluons which are similar to the gauge rotations we saw earlier in light cone gauge. Moreover, the phase factors also guarantee that \fig{fig:cgthree} is the only non-vanishing diagram involving two nucleons from each nucleus. For example, with the same approximation used in obtaining \eq{eq:covcal}, diagram \fig{fig:cgzero} gives
\begin{equation}
  \begin{split}
  &\int dp_{1+}dk_{1-}\frac{e^{ip_{1+}(x_{1-}-x_{2-})}e^{ik_{1-}(x'_{1+}-x'_{2+})}}{\big[(p_{1}+k_{1})^2+i\epsilon\big]\big[(l-k_{1}-p_{1})^2+i\epsilon\big]}\\
  &=\int dp_{1+}\int dk_{1-}\frac{e^{ip_{1+}(x_{1-}-x_{2-})}}{\big[2p_{1+}k_{1-}-(\ul{k}_{1}+\ul{p}_{1})^2+i\epsilon\big]}\frac{e^{ik_{1-}(x'_{1+}-x'_{2+})}}{\big[2l_{+}(l_{-}-k_{1-})-(\ul{l}-\ul{k}_{1}-\ul{p}_{1})^2+i\epsilon\big]}=0.
\end{split}
\end{equation}
Since the $k_{1-}$-pole in the second propagator lies opposite to the direction of the contour distortion indicated by the phase factor $e^{ik_{1-}(x'_{1+}-x'_{2+})}$. Such diagrams give zero as required by causality in covariant gauge.
\begin{figure}[h]
  \centering
  \includegraphics[width=10cm]{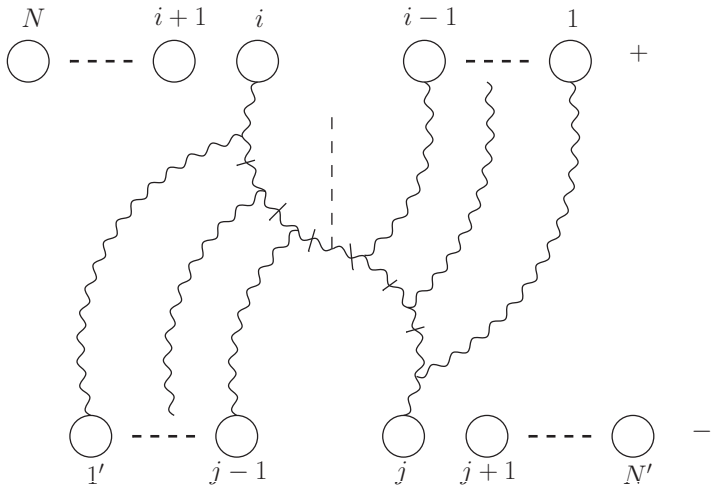}
  \caption{In nucleus-nucleus collisions the hard gluons are multiply scattered and their transverse momenta are broadened. Gluons from nucleons $i$ and $j$ are effectively put on-shell after each scattering.}
  \label{fig:covariantgauge}
\end{figure}

It is straightforward to generalize the calculation to an arbitrary number of nucleons as shown in  \fig{fig:covariantgauge}. Suppose that the hard gluons are coming from $i$th and $j$th nucleons of nucleus 1 and 2 respectively, the gluon from the $i$th nucleon in nucleus 1 can be scattered by the nucleons coming before the $j$th nucleon in nucleus 2. Similarly, the gluon which comes from the $j$th nucleon in nucleus 2 can be scattered by the nucleons coming before the $i$th nucleon in nucleus 1. One or two gluons exchanges can happen as the gluon passes  each nucleon, and after each scattering the produced gluon is put on shell by a contour distortion. We see that a hard gluons has to travel through a certain length in the nuclear matter and its transverse momentum is broadened by the multiple scattering. This process can be described by a classical diffusion equation in momentum space \cite{diffusion} and the transverse momentum distribution can be found by solving this equation. The solution of gluon diffusing in nuclear matter is found in \cite{pA}. We will use the results derived in those two papers. The unintegrated gluon distribution at the final stage of multiple scattering can be written as
\begin{equation}
  \tilde{N}(\ul{x})=\int d^2bdz_{0}\,\rho_{0}\tilde{N}_{0}(\ul{x})\tilde{f}(z,\ul{x})\bigg|_{z=z_{0}},
\label{eq:ndiffusion}
\end{equation}
where $\rho_{0}$ is the nuclear density and is assumed to be uniform throughout the nucleus, $\tilde{N}_0(\ul{x})$ is the initial gluon distribution which can be taken to be $xG(x,1/\ul{x}^2)$, $\tilde{f}(z,\ul{x})$ is the probability distribution for the gluon to have transverse coordinate $\ul{x}$ at a longitudinal position $z$, $z_{0}$ is the longitudinal position where the hard gluon finally emerges from the nuclear medium and is ready to produce the scalar particle. For the gluon from $i$th nucleon we can take $z_{0}\approx z_{j-1}$. In the case of gluon $p_{t}$-broadening, for a spherical nucleus, $\tilde{f}(z,\ul{x})$ can be written as
\begin{equation}
  \tilde{f}(z,\ul{x})=\exp\bigg\{-\frac{z+\sqrt{R^2-b^2}}{8\sqrt{R^2-b^2}}\ul{x}^2Q_{s}^{2}\bigg\},
\end{equation}
where $R$ is the radius of the nucleus and $b$ is the impact parameter. Since the hard gluon can come from any nucleon in the nucleus, we have to sum up all the possibilities which corresponds to the $dz_{0}$ integration in \eq{eq:ndiffusion}. It gives us
\begin{equation}
  \begin{split}
  \tilde{N}_{1}(\ul{x})&\propto\int^{\sqrt{R^2-b^2}}_{-\sqrt{R^2-b^2}}dz_{0}\, \exp\bigg\{-\frac{z_{0}+\sqrt{R^2-b^2}}{8\sqrt{R^2-b^2}}\ul{x}^2Q_{s}^{2}\bigg\}\\
  &\propto\frac{8\sqrt{R^2-b^2}}{\ul{x}^2Q^{2}_{s}}\big(1-e^{-\ul{x}^2Q_{s}^{2}/4}\big),
\end{split}
\end{equation}
which is the WW type gluon distribution and is the same as \eq{eq:unintww}. This one-to-one correspondence between the LC calculation and the covariant calculation is also explained in \cite{pA}. It is now quite convincing that WW gluon distribution is the right gluon distribution that should be used for this process. It is interesting to note that in covariant gauge, as in light cone gauge, the nucleons ``behind'' the nucleon which gives the hard gluon creating the scalar particle are viewed as non-interacting, which is the very property that gives rise to the WW gluon distribution. However, in covariant gauge, the diagrams do not clearly indicate that the cross section is factorizable.

\section*{Acknowledgments} 
The author would like to thank A.~H.~Mueller for suggesting this work and many stimulating and informative discussions, as well as for reading the manuscript. Partial support is provided by the U.S. Department of Energy.

\appendix
\appendixpage
\section{Other choices of $i\epsilon$'s}
\begin{figure}[h]
  \centering
  \subfigure[]{\label{fig:a}
    \includegraphics[width=5cm]{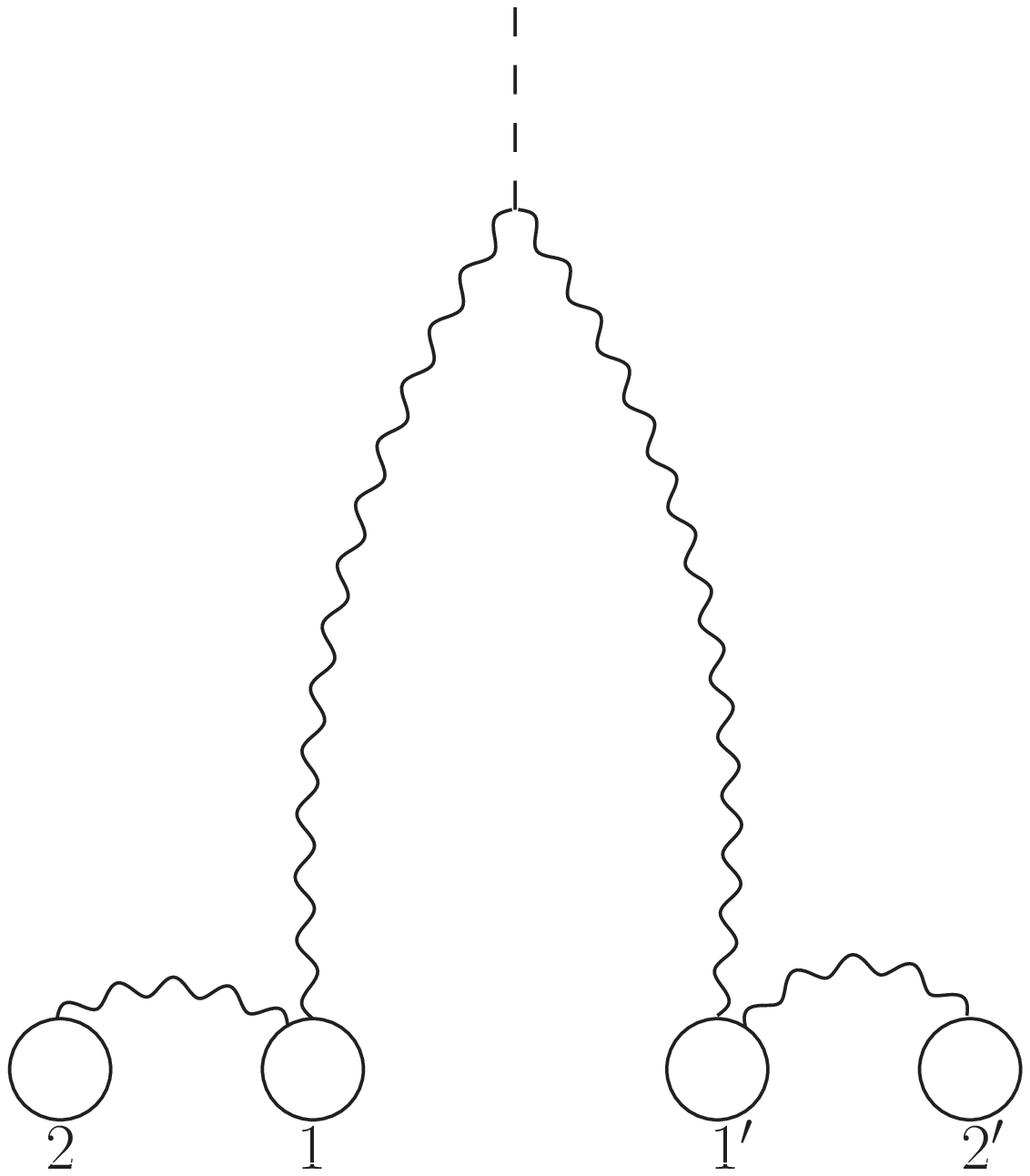}}
  \subfigure[]{\label{fig:b}
    \includegraphics[width=5cm]{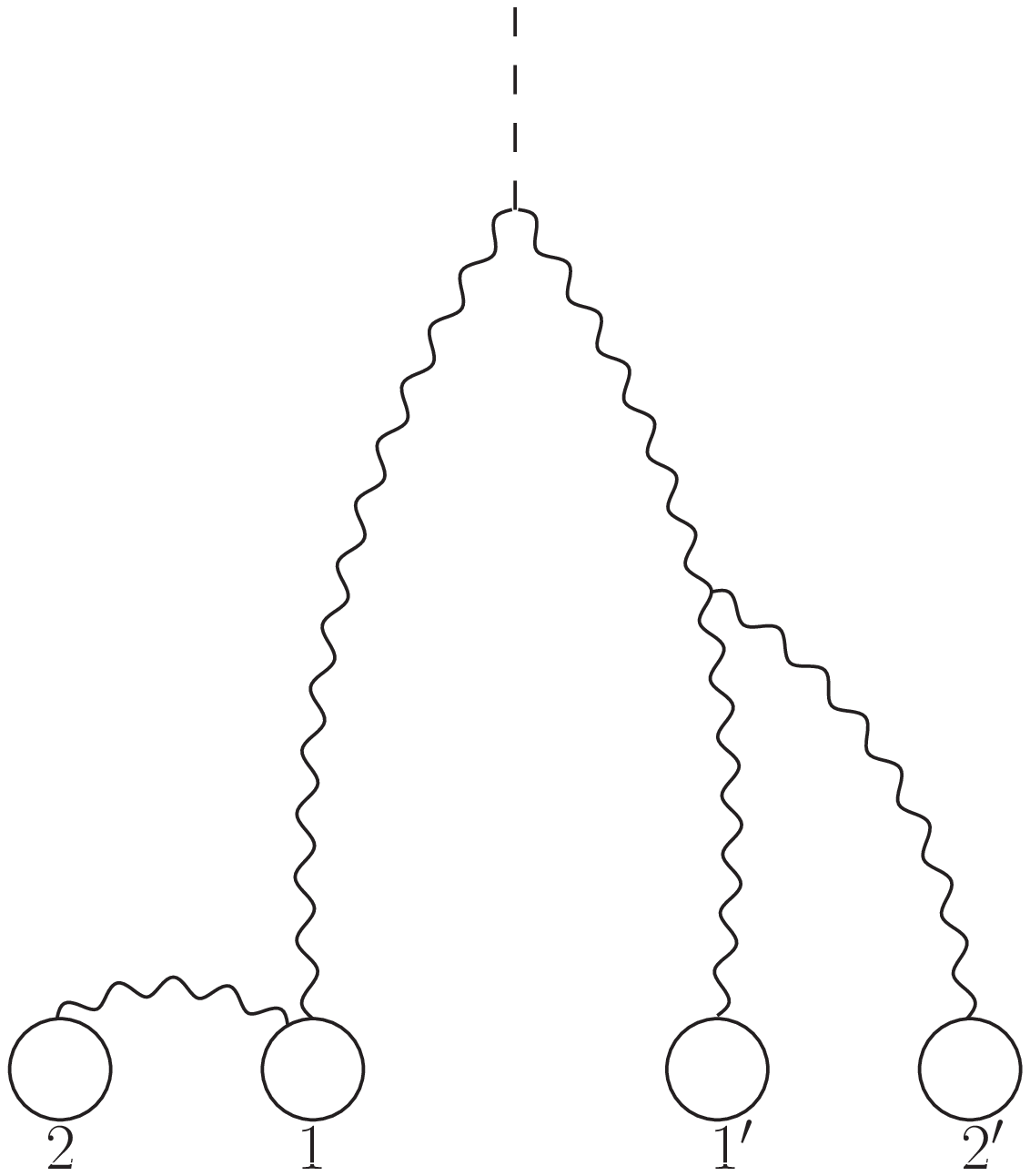}}
 \subfigure[]{\label{fig:c}
    \includegraphics[width=5cm]{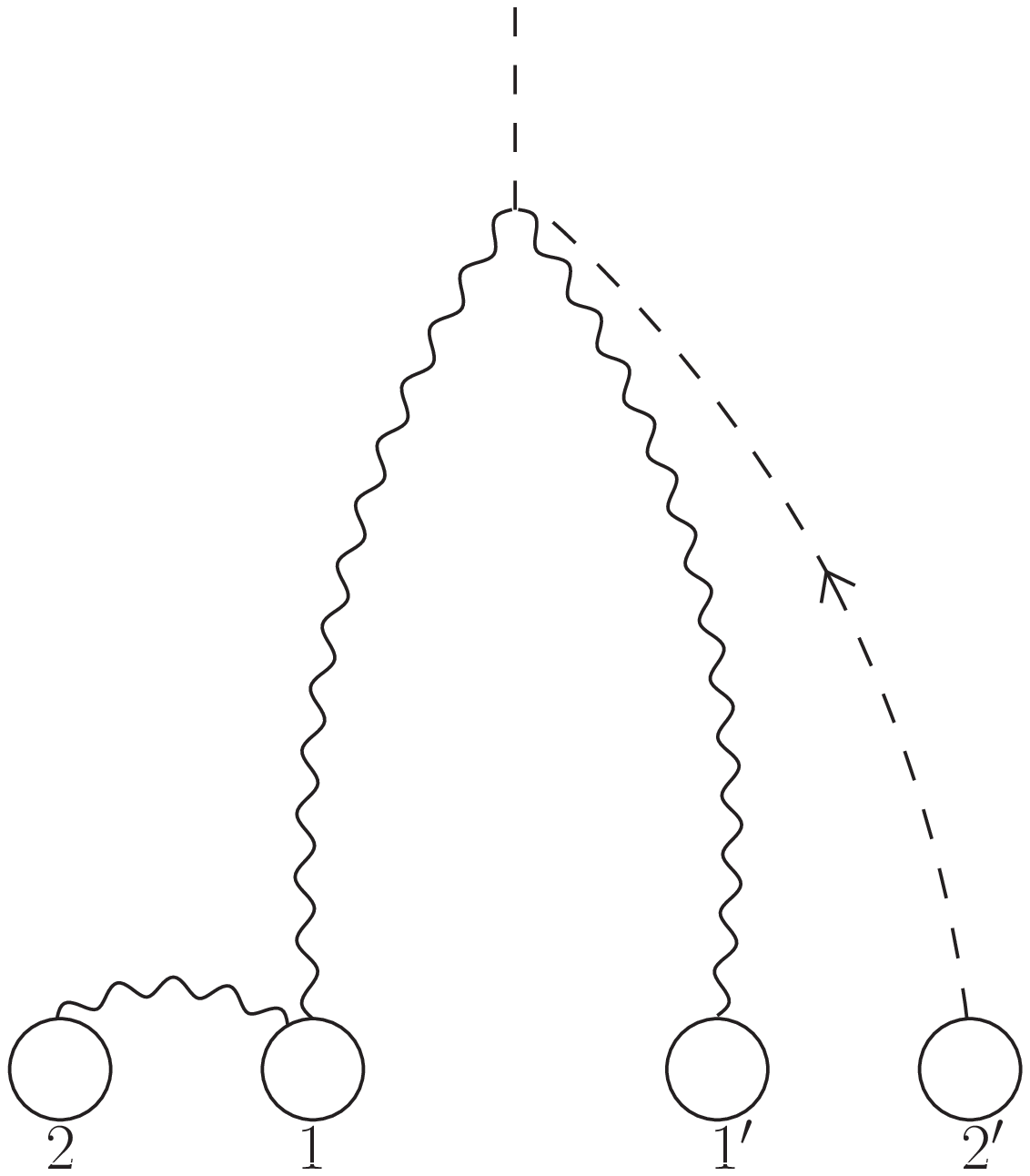}}
 \subfigure[]{\label{fig:d}
    \includegraphics[width=5cm]{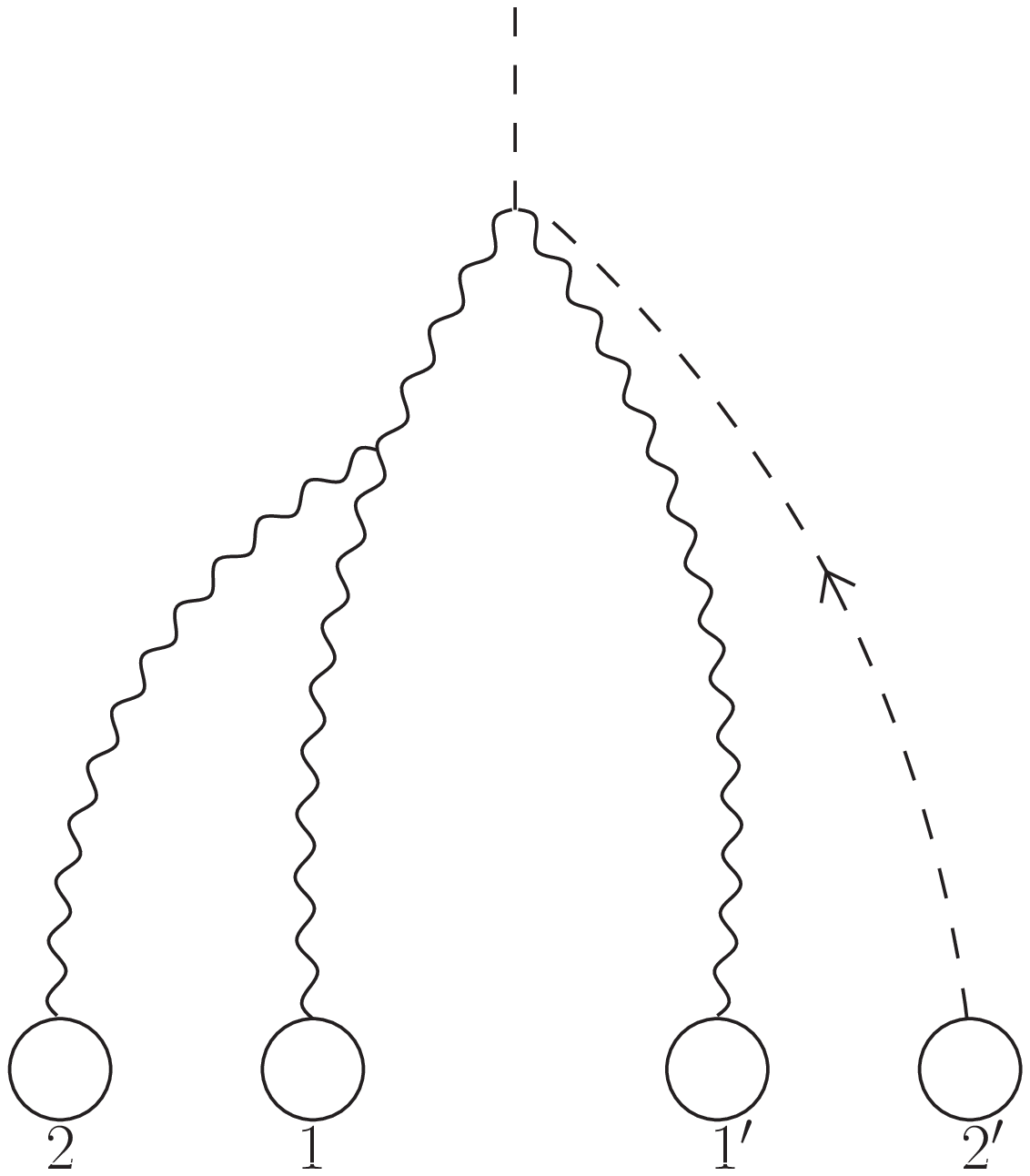}}
 \subfigure[]{\label{fig:e}
    \includegraphics[width=5cm]{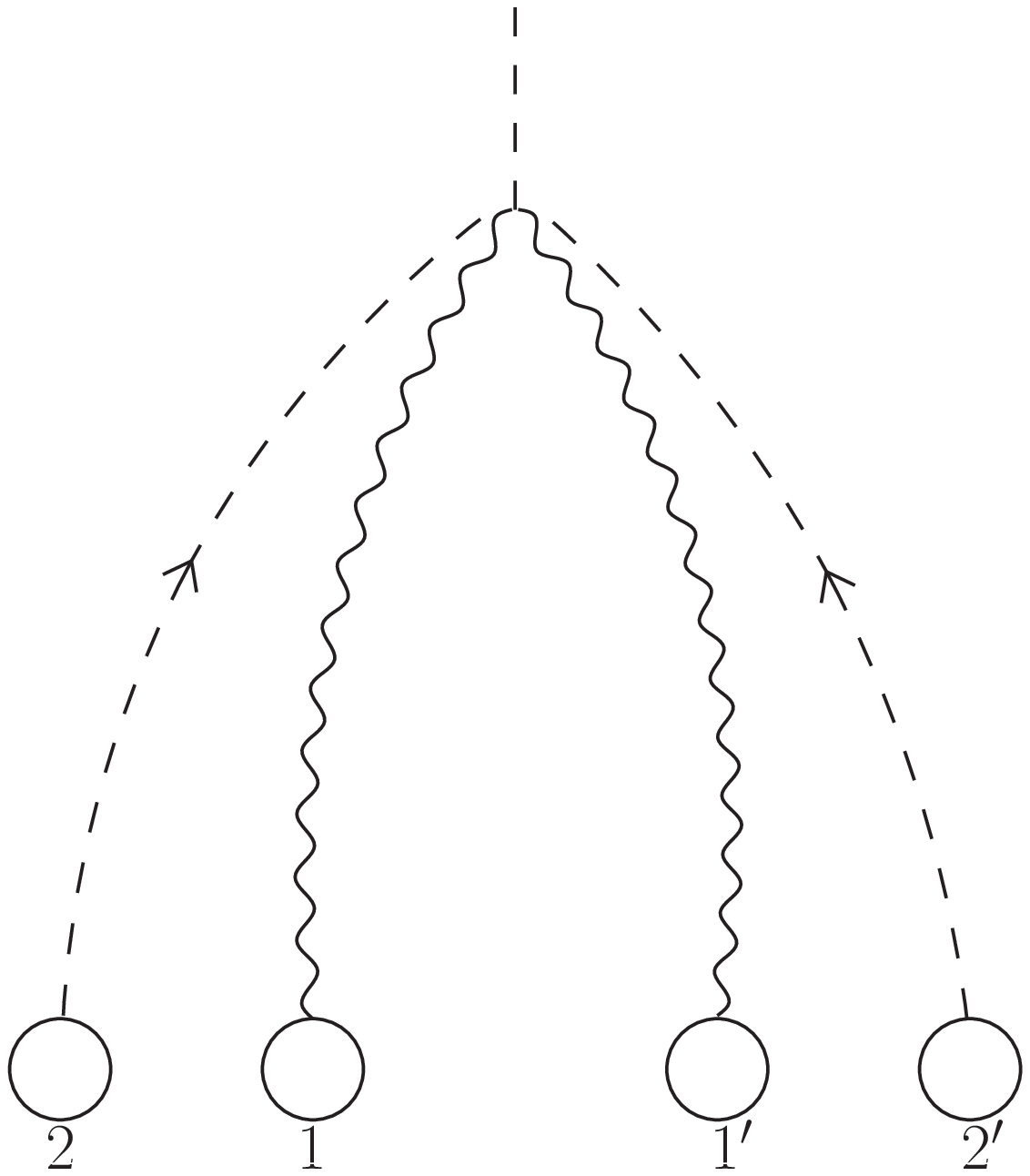}}
\caption{Collisions are viewed in a different choice of $i\epsilon$'s. The hard gluons come from nucleons 1 and $1'$. The soft gluons from nucleons 2 and $2'$ become the gauge rotations.}
\label{fig:gaugeone}
\end{figure}

In this appendix, we will show how a different regularization in the LC propagator gives the same result. Although there are three more cases in addition to the one discussed in the body of the paper, here we will only study the most complicated one where the gluons in the nuclei propagate in the negative $x_{-}$($x_{+}$)-direction in $A_{+}=0$($A_{-}=0$) gauge. The two remaining cases can also be shown to give the same result by applying the same technique. Now the choices of $i\epsilon$'s are taken to be $(k_{+}-i\epsilon)$ and $(k_{-}-i\epsilon)$ for $A_{+}=0$ and $A_{-}=0$ gauge respectively, which means that gluons in $A_{+}=0$($A_{-}=0$) gauge propagate from a large $x_{-}$($x_{+}$)-coordinate to a small $x_{-}$($x_{+}$)-coordinate. With all the gluons propagating in this way initial state interactions are complex and entanglement between the soft gluons can occur. Even nucleons belonging to different nuclei could affect each other, therefore the factorizability is not clear at the very beginning. This is quite different from what we have seen previously where there is no initial state interactions between the two nuclei. We will see that the STW identities guarantee that the cross section is still factorizable and is the same as the one obtained previously. 

\begin{figure}[h]
\subfigure[]{\label{fig:f}
    \includegraphics[width=5cm]{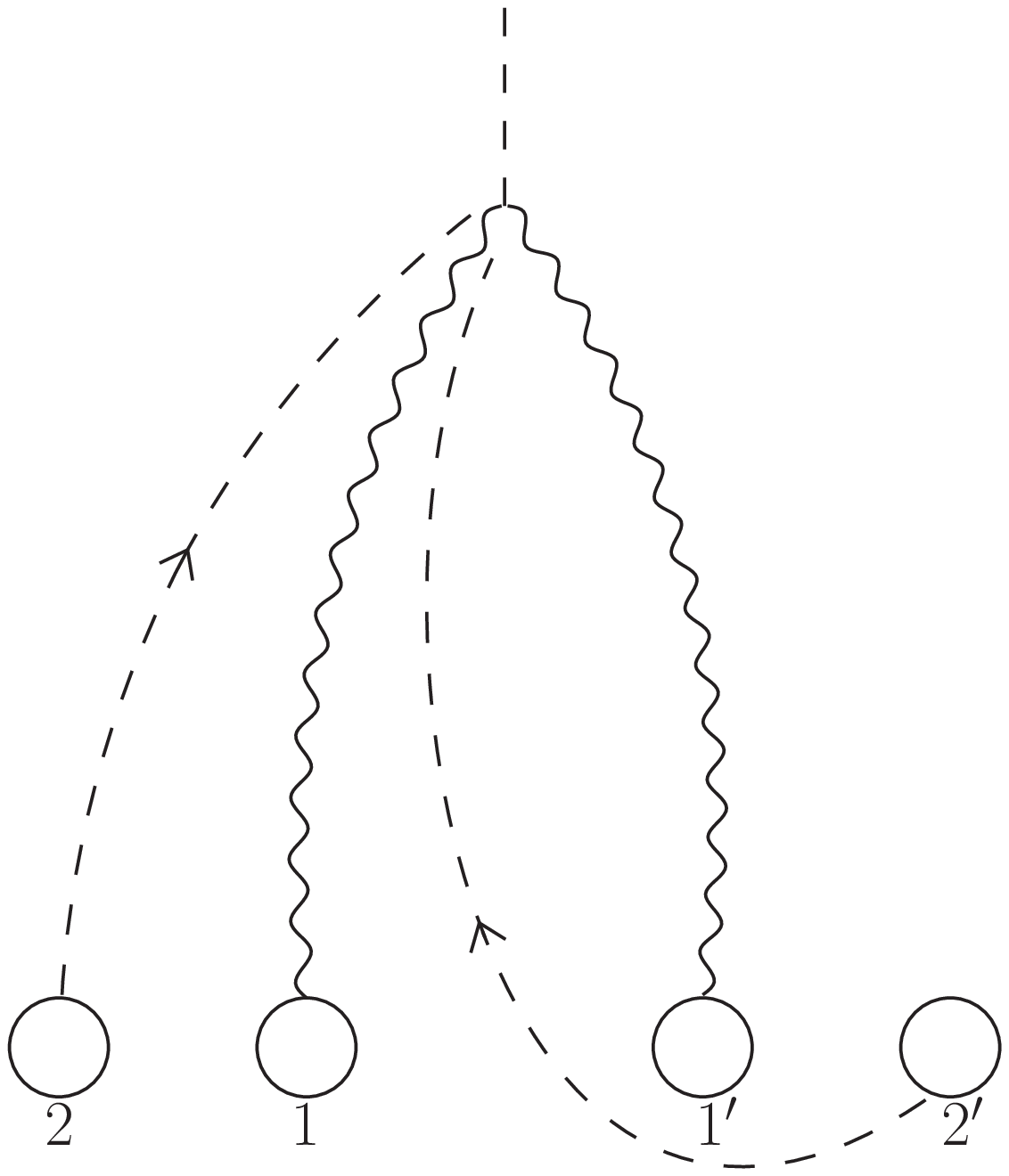}}
 \subfigure[]{\label{fig:m}
    \includegraphics[width=5cm]{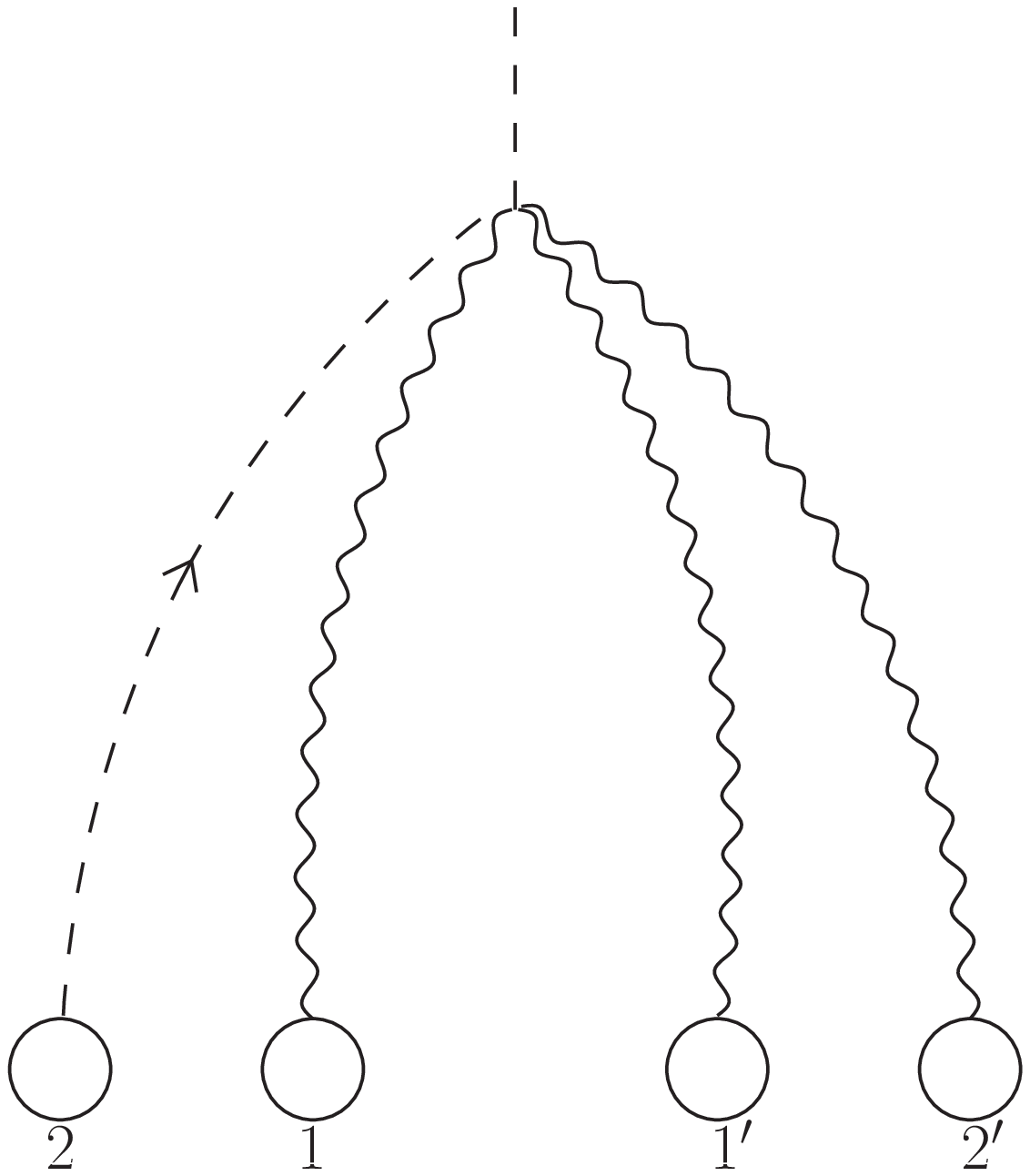}}
 \subfigure[]{\label{fig:n}
    \includegraphics[width=5cm]{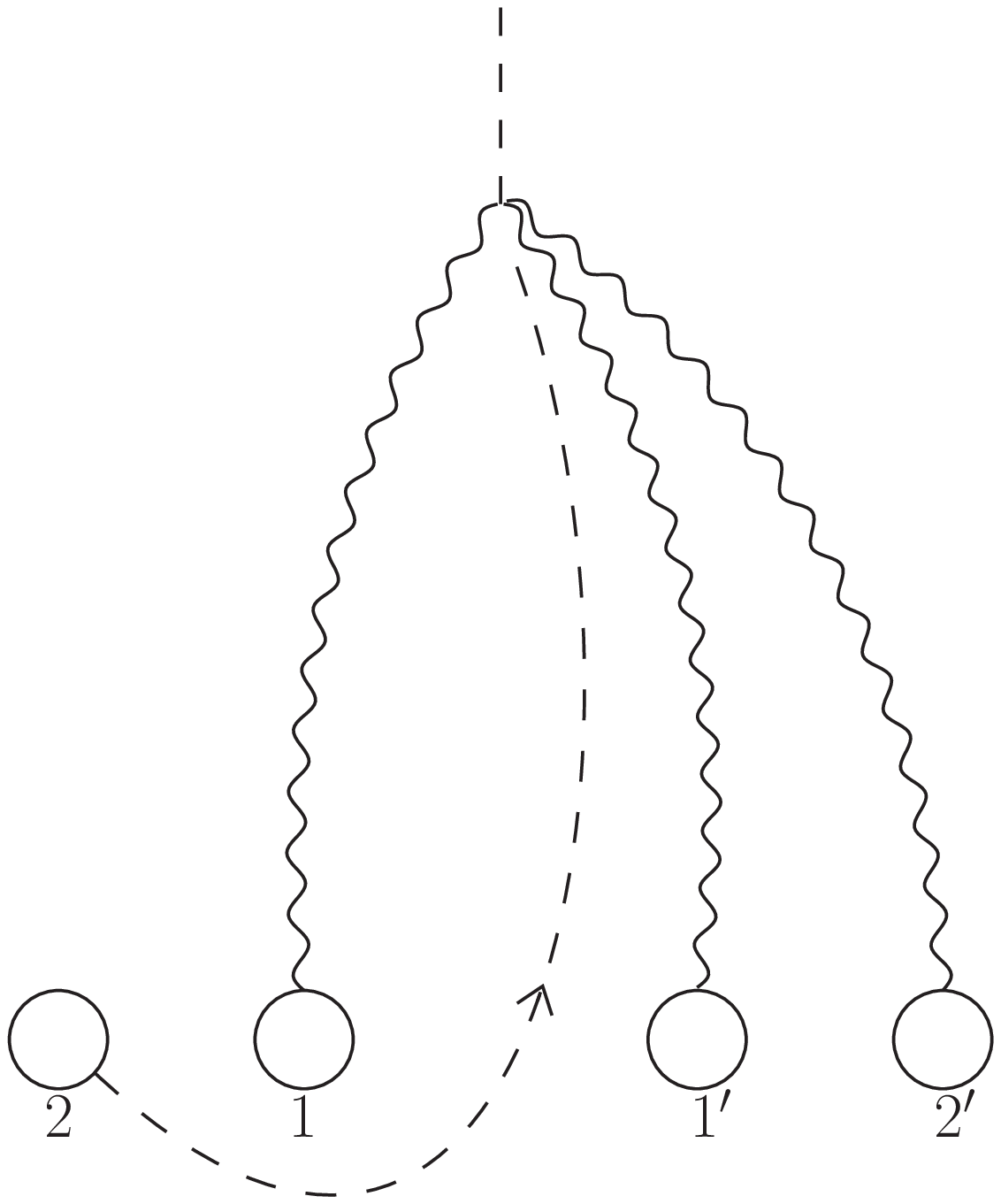}}
 \subfigure[]{\label{fig:g}
    \includegraphics[width=5cm]{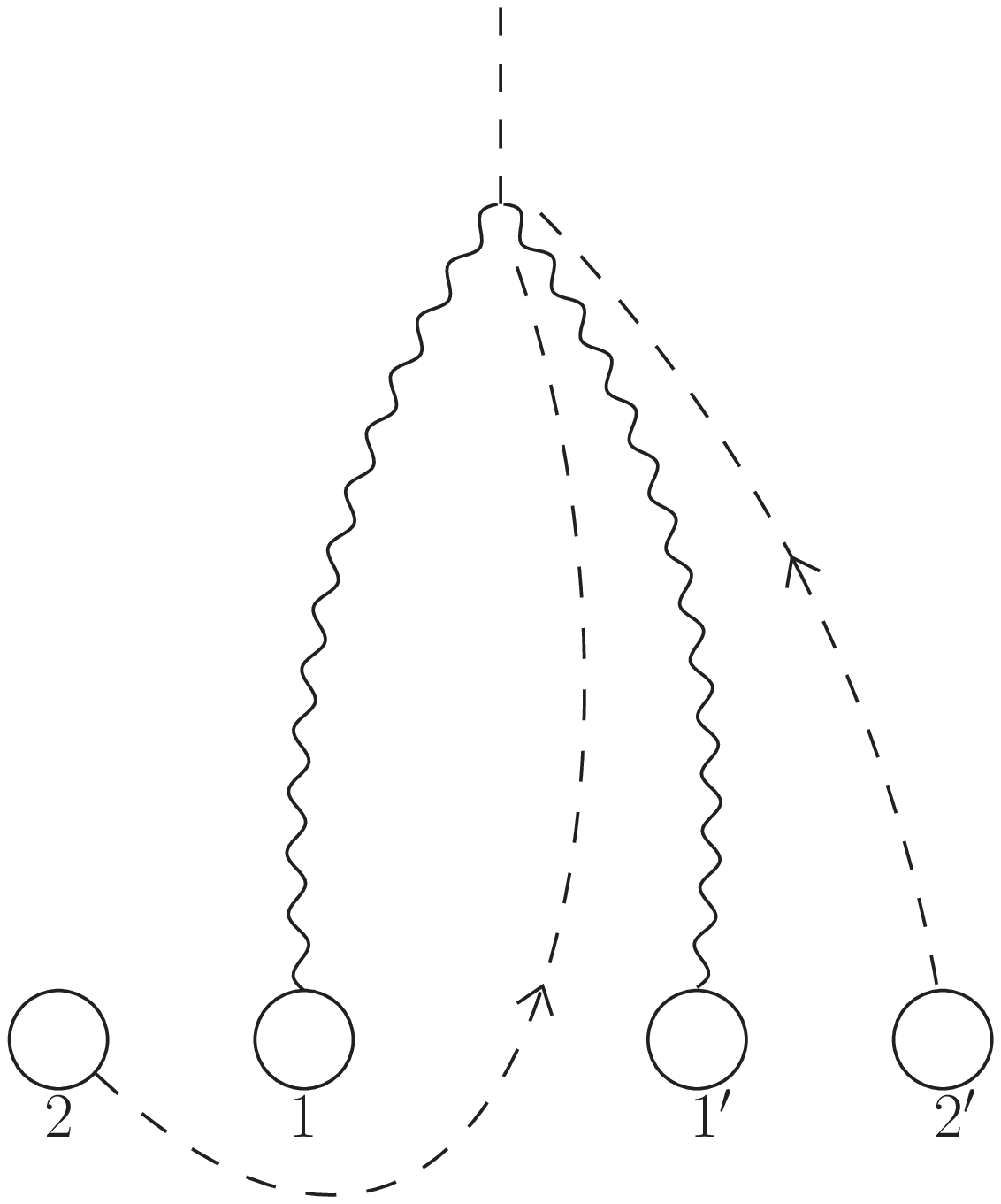}}
 \subfigure[]{\label{fig:h}
    \includegraphics[width=5cm]{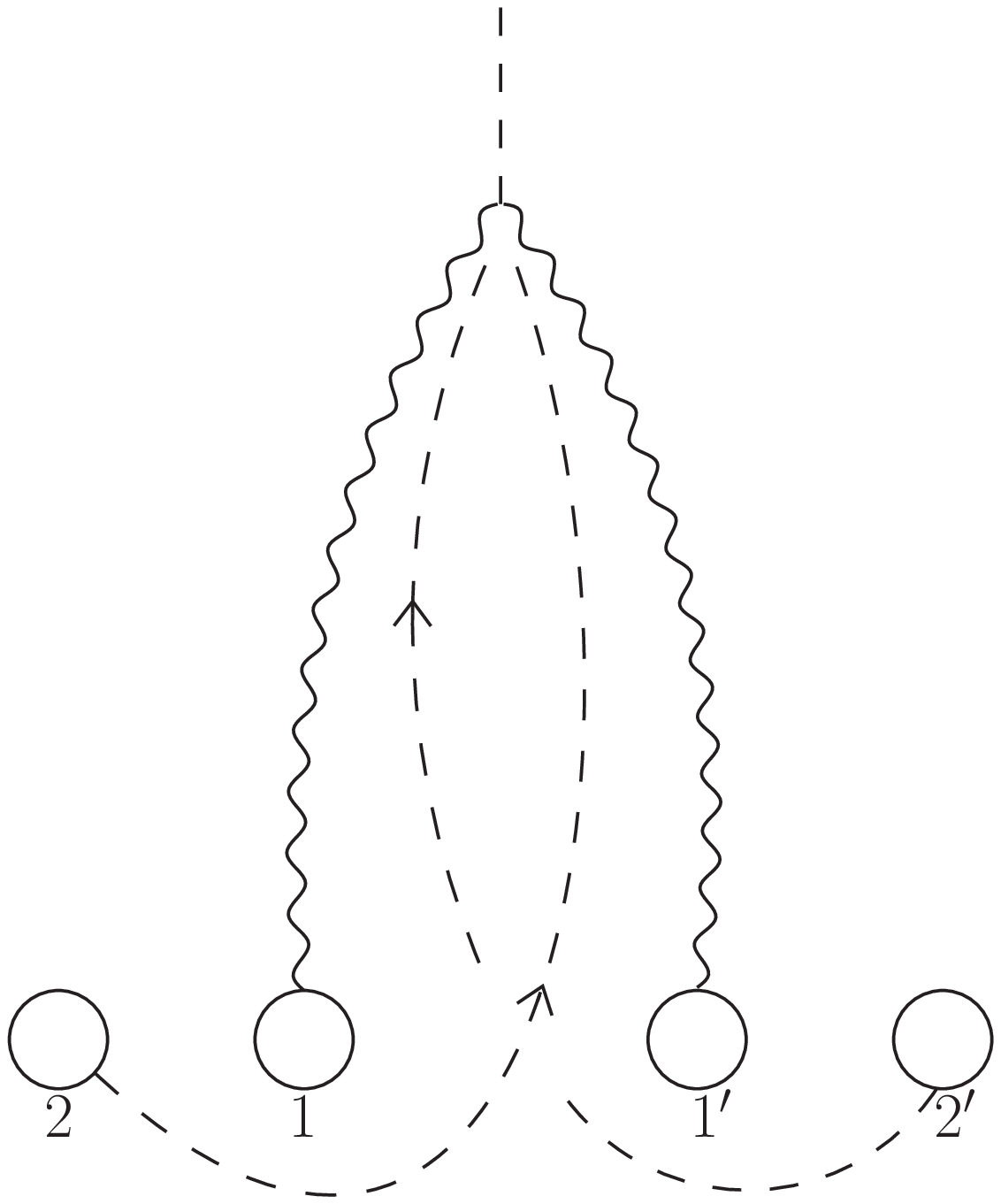}}
\caption{Additional diagrams that are needed in the case where the hard gluons are from nucleons 1 and $1'$.}
\label{fig:gaugefirst}
\end{figure}
In light cone gauge we have the privilege to specify where the hard gluons come from, so in order to simplify the calculation we further classify the diagrams by how the hard gluons are connected to the nucleons and group them into several distinct classes. The advantage of doing so is that, on one hand, one can avoid double counting diagrams and, on the other hand, the gluons that we have to add to the diagrams should be soft so that we can apply the STW identities to them. 

We start with the case where the hard gluons are coming from the first nucleons on each side as shown in \fig{fig:gaugeone}. There are still two more soft gluon lines that should be added to the graph. In order to avoid double counting we can fix one of the soft lines first, for example, the gluon line connecting nucleons 1 and 2 in \fig{fig:a} and \fig{fig:b}. Then draw different connections of the second soft line so that we can apply the STW identities. If we add \fig{fig:a} and \fig{fig:b} together the soft gluon from nucleon $2'$ becomes the gauge rotation shown in \fig{fig:c}. Then we focus on the first soft line and enumerate its possible connections, now as a dashed gauge line, while doing the same thing to the second soft gluon. We obtain another diagram shown in \fig{fig:d}. Adding \fig{fig:c} and \fig{fig:d} together, we change the soft gluon from nucleon 2 to a gauge rotation and obtain the diagram shown in \fig{fig:e}. Since the gauge rotation acts independently on the hard gluons, the gauge rotation from nucleon $2'$ can also attach to the hard gluon from nucleon 1. One obtains an additional diagram shown in  \fig{fig:f}. We need one more diagram, \fig{fig:m}, in order to apply the STW identities. But such a diagram with a three-gluon-scalar vertex is suppressed by a power of mass squared of the scalar particle $M^{2}=l^2$ and can be neglected in our approximation. It is shown here only for the sake of completeness. If one add all the diagrams \fig{fig:e}, \fig{fig:f}, \fig{fig:m} together, the STW identities guarantee that they give zero contribution. A similar argument can also be made to obtain the diagrams shown in \fig{fig:n}, \fig{fig:g} and \fig{fig:h}, their sum is also zero. Therefore the soft gluons from the last interacting nucleons do not contribute to the collisions in this specific choices of $i\epsilon$'s. This result is what one should expect from the STW identities. Since the identities tell us that if we enumerate all possible insertions of a longitudinal polarized gluon line to a certain graph their sum should be zero. This is exactly what we have in this case, where the gluon fields coming from either nucleon 2 or $2'$, the last interacting nucleon, can gauge rotate everything that comes before them. Therefore all possible connections should add up to zero. So it is not surprising that even though we might have a huge number of diagrams, gauge invariance guarantees that many of them add to give no contribution.

\begin{figure}[h]
  \centering
  \subfigure[]{\label{fig:i}
    \includegraphics[width=5cm]{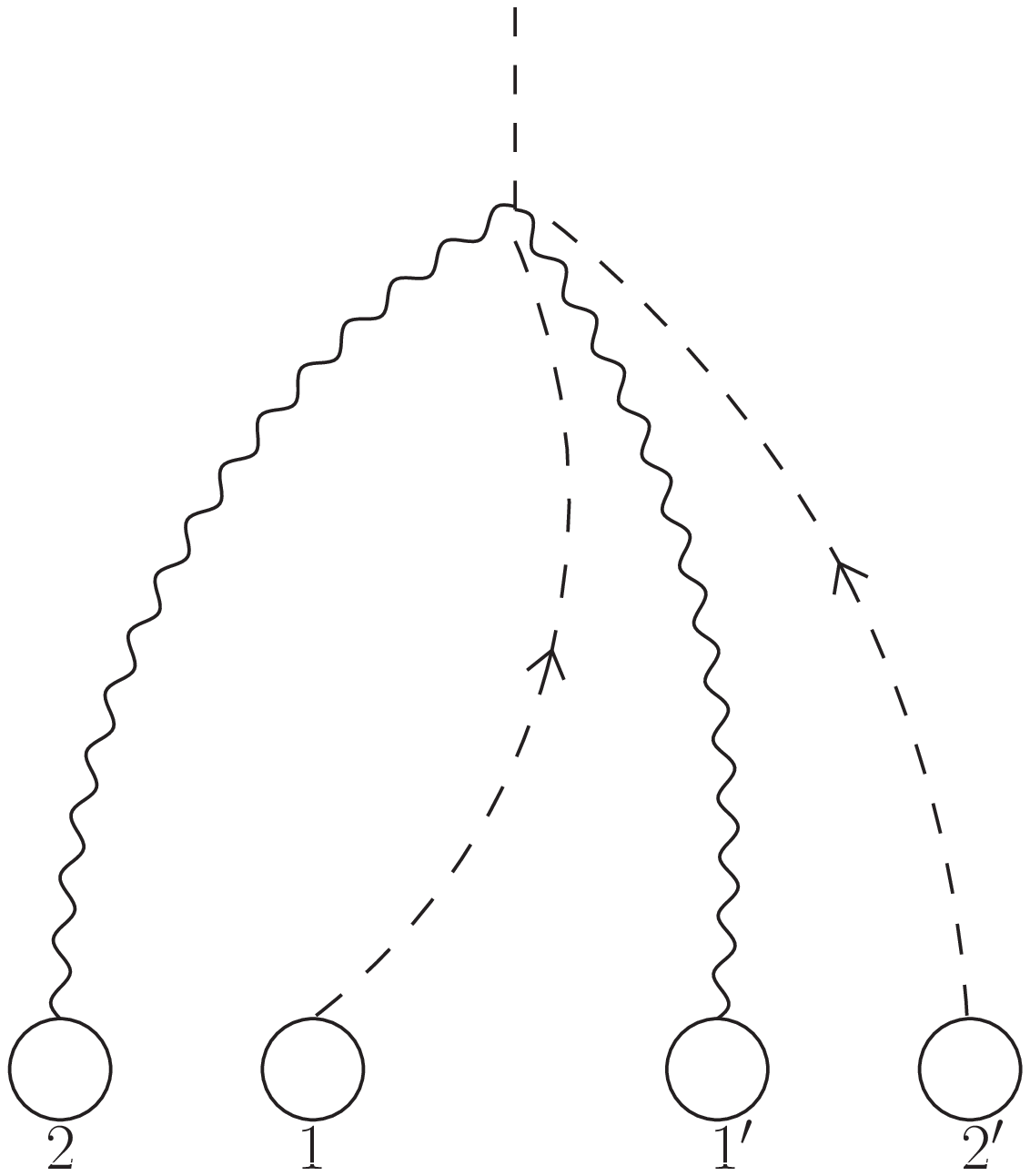}}
  \subfigure[]{\label{fig:j}
    \includegraphics[width=5cm]{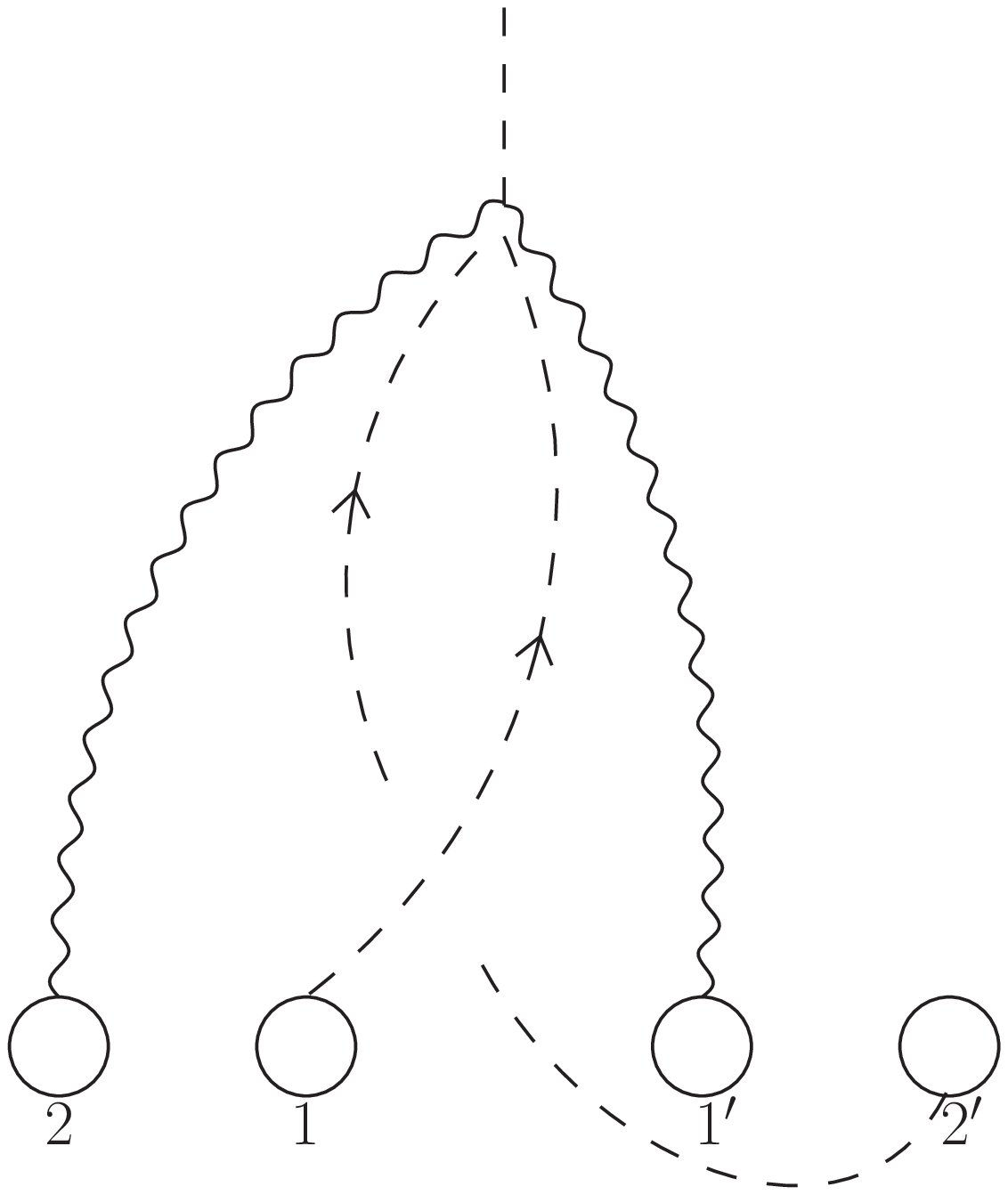}}
 \subfigure[]{\label{fig:l}
    \includegraphics[width=5cm]{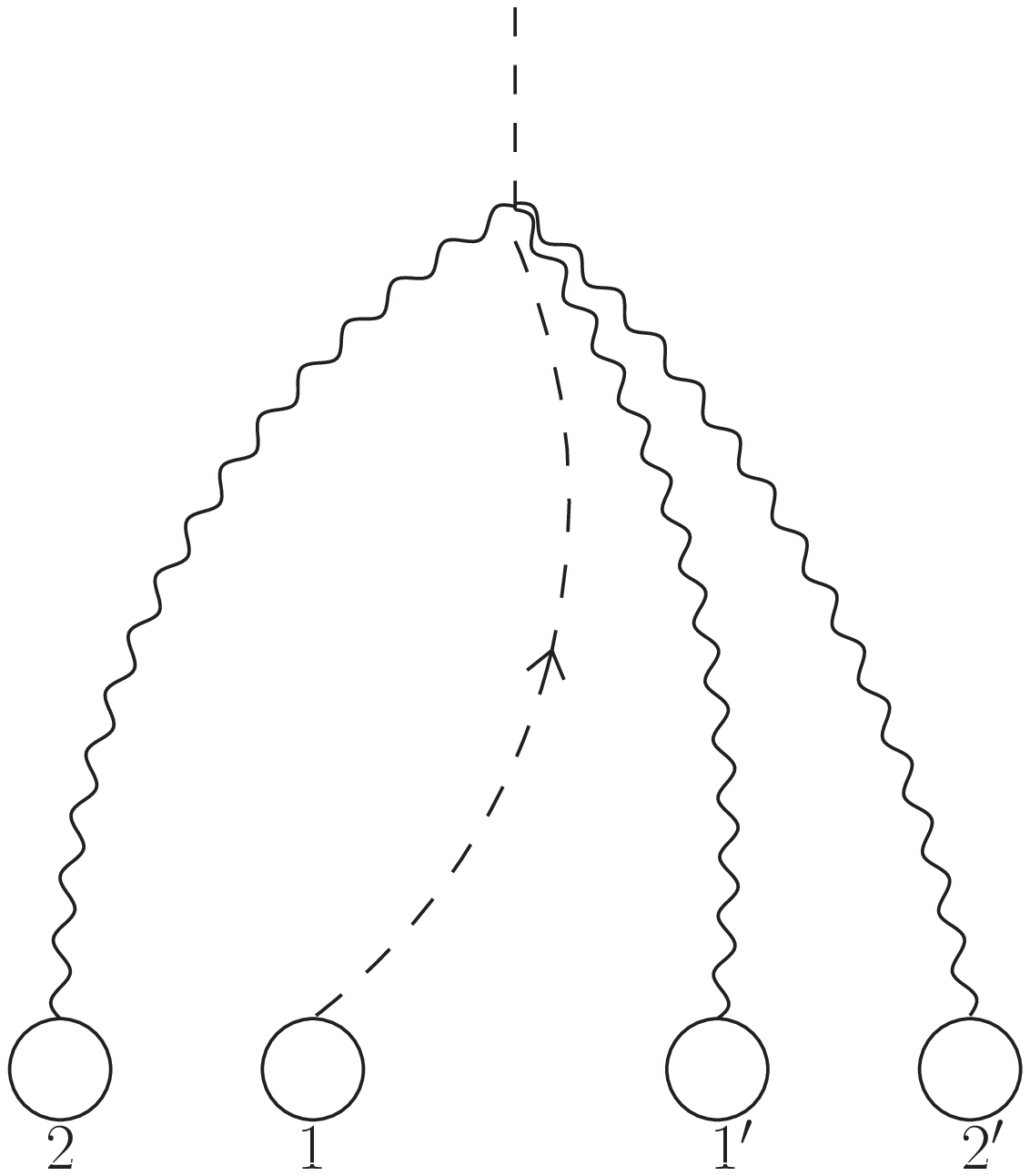}}
  \subfigure[]{\label{fig:o}
    \includegraphics[width=5cm]{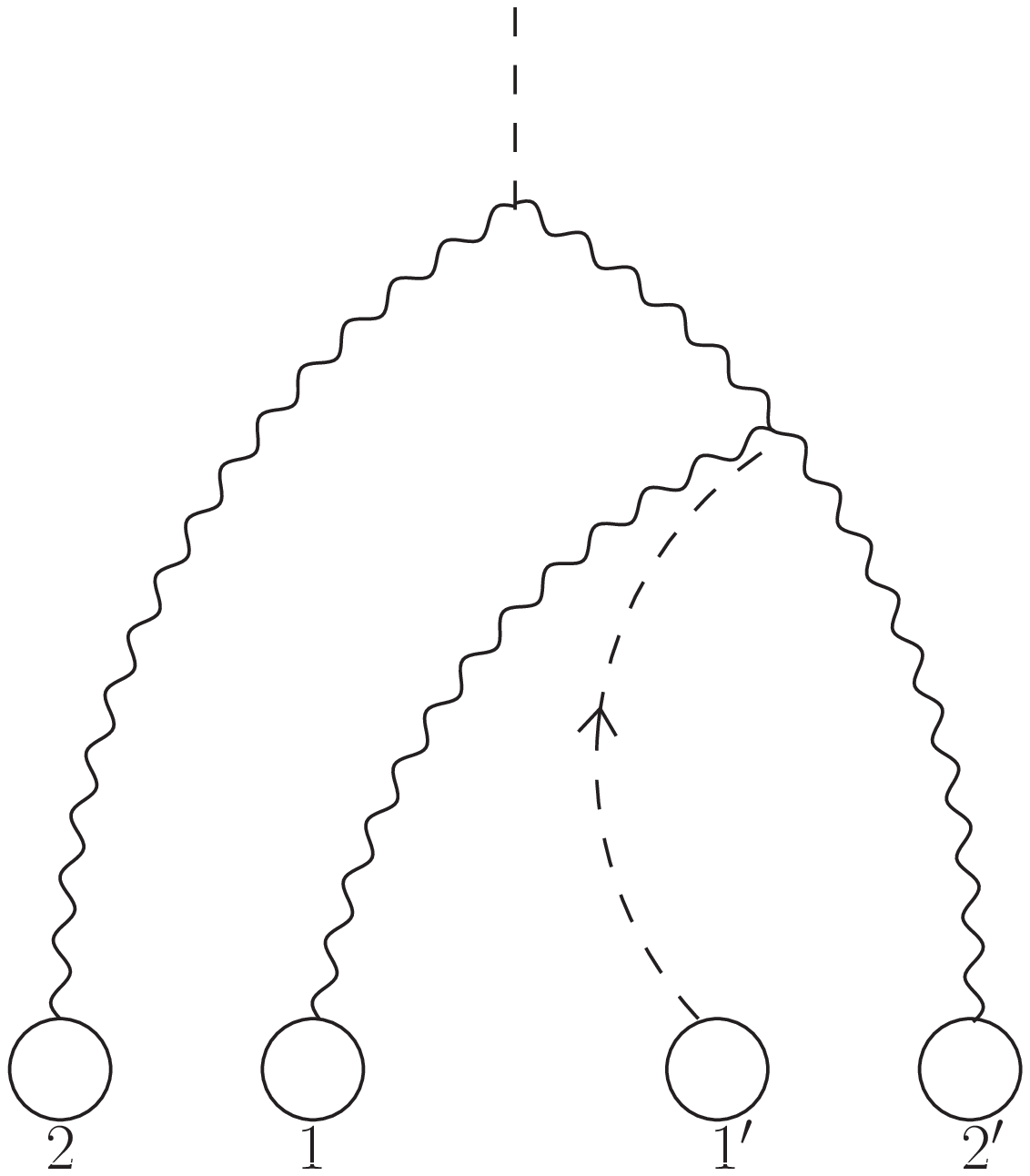}}
  \subfigure[]{\label{fig:p}
    \includegraphics[width=5cm]{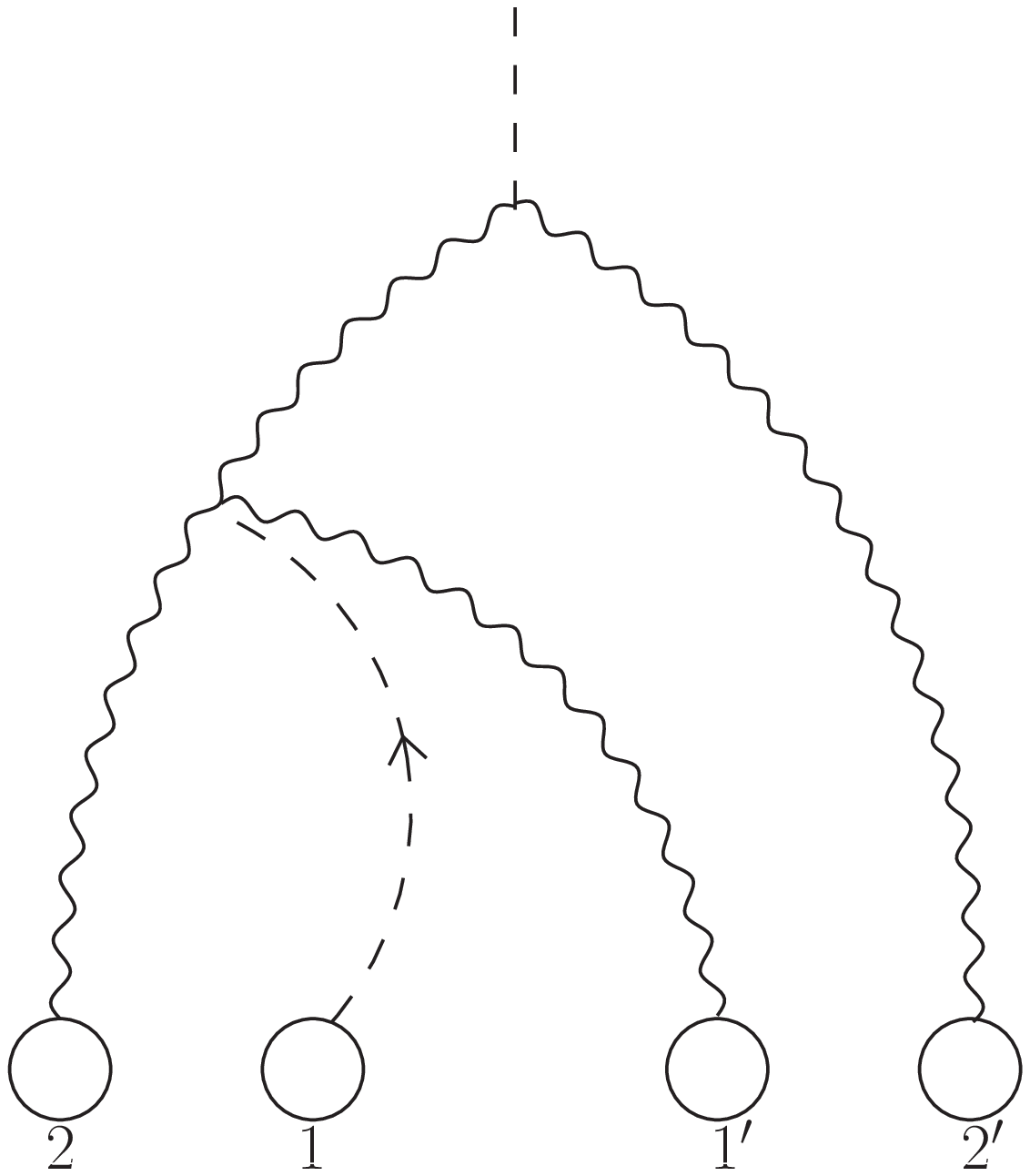}}
  \subfigure[]{\label{fig:k}
    \includegraphics[width=5cm]{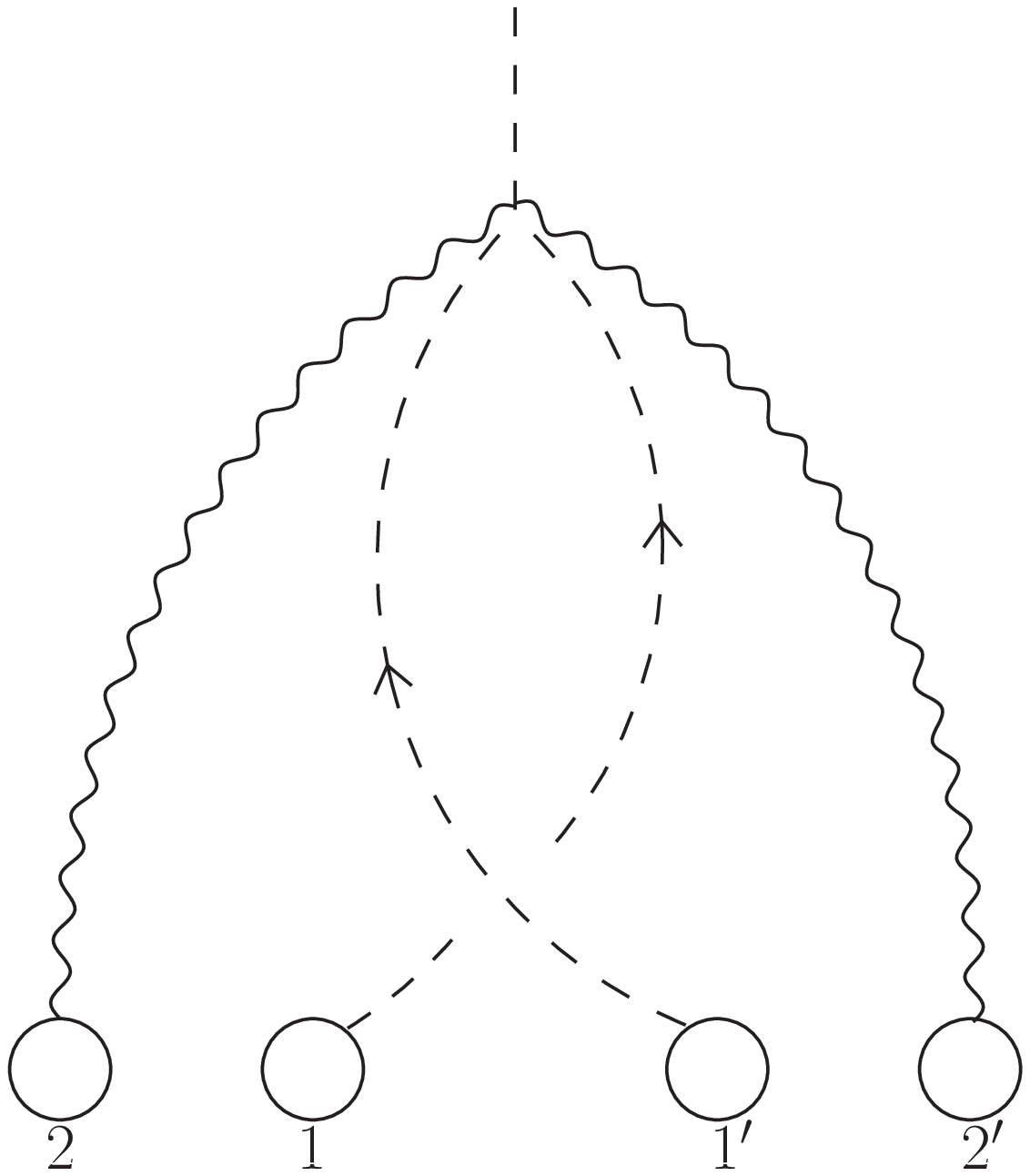}}
\caption{Other two different cases where the hard gluons come from nucleons 2 and $1'$ and nucleons 2 and $2'$. The later one gives the dominant contribution.}
\label{fig:other}
\end{figure}

Similarly, one can also consider the case where the hard gluons are coming from nucleons 2 and $1'$, shown in \fig{fig:i}, \fig{fig:j} and \fig{fig:l}, and the symmetric case, where the hard lines are coming from nucleons $2'$ and 1, which is not shown in the figure. Since these two cases are essentially the same, it is sufficient to study only one of them. Due to the STW identities, they also give us no contribution.

Finally, the last case is the one where the hard lines are coming from nucleons 2 and $2'$. Since now the soft gluons can gauge rotate only certain parts of the diagram, it should give a nonzero result. If there is an entanglement between the two soft gluons, one obtains diagrams like \fig{fig:o} and \fig{fig:p}. However, this kind of diagrams are suppressed by $M^2$. Since the three-gluon vertex brings in an additional factor of transverse-momentum squared, a factor of $M^2$ comes from the denominator to make up the right dimension. Therefore those diagrams have small contributions in our approximation. Moreover, one could also make a contour distortion explained in the appendix to make such diagrams zero. If the two soft gluons are directly connected to the hard gluons, one obtains \fig{fig:k} which is the dominant contribution in this choice of $i\epsilon$'s. One can immediately recognize that this diagram is exactly the same as the one in \fig{fig:gauge}. 

\section{Suppression of the initial state interactions}
\begin{figure}[h]
  \centering
  \includegraphics[width=8cm]{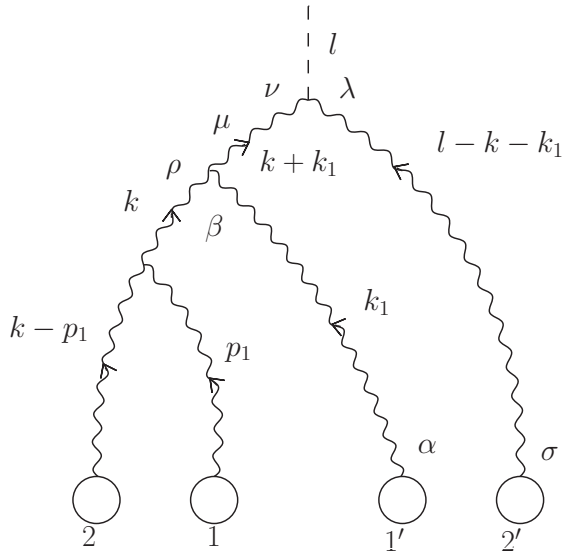}
  \caption{The initial state interactions are suppressed by the heavy mass of the scalar particle.}
  \label{fig:initial}
\end{figure}
In Section \ref{sec:calculation} we argued that the initial state interactions are suppressed by the scalar mass $M$ in the $(k_{1+}+i\epsilon)$ and $(k_{-}+i\epsilon)$ gauge choice. Here we will do an explicit calculation to show that it is true. We take the example mentioned in Section \ref{sec:calculation}. The diagram is shown in \fig{fig:initial}, which is quite similar to \fig{fig:gaugerotation:a} but the connection of $k_{1}$-line is changed. Let us focus on the part of the diagram that involves the gluons from nucleus 2, nucleons $1'$ and $2'$. The relevant factors in the integrand are
\begin{equation}
\begin{split}
  &\frac{i}{k_{1}^2+i\epsilon}\frac{\bar{\eta}_{\alpha}k_{1\beta}^{\perp}}{k_{1-}+i\epsilon}\Gamma_{\rho\beta\mu}\frac{i}{(k+k_{1})^{2}+i\epsilon}\bigg[g_{\mu\nu}-\frac{\eta_{\mu}(k+k_{1})_{\nu}}{(k+k_{1})_{+}+i\epsilon}-\frac{\eta_{\nu}(k+k_{1})_{\mu}}{(k+k_{1})_{+}-i\epsilon}\bigg]\\
  &\times v_{\nu\lambda}\frac{i}{(l-k-k_{1})^{2}+i\epsilon}\frac{\bar{\eta}_{\sigma}(l-k-k_{1})^{\perp}_{\lambda}}{(l-k-k_{1})_{-}+i\epsilon},
\end{split}
\label{eq:suppression}
\end{equation}
where $\bar{\eta}\cdot k=k_{-}$. Since the big momentum component that flows through the $(k+k_{1})$-propagator is the big plus momentum component coming from nucleon 2, it is natural to use $A_{+}=0$ gauge for the $(k+k_{1})$-propagator. We can distort the contour to the upper half $k_{1-}$-plane to pick up the $k_{1-}=(l-k)_{-}+i\epsilon$ pole. Since $p_{1-}\approx 0$ and $(k-p_{1})_{-}\approx 0$, then $(l-k)_{-}\approx l_{-}\sim M$ is large in the center of mass system of the colliding nuclei. In order to compensate for the $k_{1-}+i\epsilon$ and $(k+k_{1})^2+i\epsilon\approx 2k_{+}k_{1-}-(\ul{k}+\ul{k}_{1})^2+i\epsilon$ denominators, we would need a $k_{1-}^{3}$ coming from the numerator, a $k_{1-}^2$ cancels the two denominators and one $k_{1-}$ is necessarily to form the large mass $M$. There are only two vertices, $\Gamma_{\rho\beta\mu}$ and $v_{\nu\lambda}$, that involve $k_{1-}$. However, the $v_{\nu\lambda}$ vertex cannot give a factor of $k_{1-}$ because of $(l-k-k_{1})_{\lambda}^{\perp}$. Therefore, there is sufficient convergence in the $k_{1-}$-plane for the contour to vanish at infinity, and such diagrams are suppressed by the mass of the scalar particle. We could arrive at an even stronger result more directly by noticing that we can replace $(l-k-k_{1})_{\lambda}^{\perp}$ by $(l-k-k_{1})_{\lambda}$ after the contour distortion and then use the current conservation in the $\lambda$-index to obtain a zero result. However, such a diagram is not suppressed in $(k_{-}-i\epsilon)$ choice for the propagators in $A_{-}=0$ gauge. This new $i\epsilon$ prescription changes the second and the last factors in \eq{eq:suppression} to $1/(k_{1-}-i\epsilon)$ and $1/[(l-k-k_{1})_{-}-i\epsilon]$, respectively. We can still distort the contour in the upper half plane to pick up the $k_{1-}=i\epsilon$ pole. All the other $k_{1-}$-poles lie on the other side of the real axis, and no suppression occurs in this situation. After the contour integration, one can replace $k_{1\beta}^{\perp}$ by $k_{1\beta}$ and then use the STW identities to simplify the diagrams.

\providecommand{\href}[2]{#2}\begingroup\raggedright

\endgroup

\end{document}